\documentclass[twocolumn]{aastex6}

\shorttitle{Influence of Jupiter on Earth's Orbital Cycles}
\shortauthors{Jonathan Horner et al.}

\begin{document}

\title{Quantifying the Influence of Jupiter on the Earth's Orbital Cycles}

\author{
  Jonathan Horner\altaffilmark{1},
  Pam Vervoort\altaffilmark{2},
  Stephen R. Kane\altaffilmark{2},
  Alma Y. Ceja\altaffilmark{2},
  David Waltham\altaffilmark{3},  
  James Gilmore\altaffilmark{4},
  Sandra Kirtland Turner\altaffilmark{2} 
}
\altaffiltext{1}{Centre for Astrophysics, University of Southern Queensland, Toowoomba, QLD 4350, Australia}
\altaffiltext{2}{Department of Earth and Planetary Sciences, University of California, Riverside, CA 92521, USA}
\altaffiltext{3}{Department of Earth Sciences, Royal Holloway University of London, Egham, UK}
\altaffiltext{4}{Australian Centre for Astrobiology, UNSW Australia, Sydney, New South Wales 2052,
Australia}
\email{jonathan.horner@usq.edu.au}

%%%%%%%%%%%%%%%%%%%%%%%%%%%%%%%%%%%%%%%%%%%%%%%%%%%%%%%%%%%

\begin{abstract}

% I think we should also include in the abstract that planetary systems with giant planets closer to the Star will result in cycles of higher frequency/shorter period, which is important for the long-term habitability of exoplanets!

A wealth of Earth-sized exoplanets will be discovered in the coming years, proving a large pool of candidates from which the targets for the search for life beyond the Solar system will be chosen. The target selection process will require the leveraging of all available information in order to maximize the robustness of the target list and make the most productive use of follow-up resources. Here, we present the results of a suite of $n$-body simulations that demonstrate the degree to which the orbital architecture of the Solar system impacts the variability of Earth's orbital elements. By varying the orbit of Jupiter and keeping the initial orbits of the other planets constant, we demonstrate how subtle changes in Solar system architecture could alter the Earth's orbital evolution -- a key factor in the Milankovitch cycles that alter the amount and distribution of solar insolation, thereby driving periodic climate change on our planet. The amplitudes and frequencies of Earth's modern orbital cycles fall in the middle of the range seen in our runs for all parameters considered -- neither unusually fast nor slow, nor large nor small. This finding runs counter to the `Rare Earth' hypothesis, which suggests that conditions on Earth are so unusual that life elsewhere is essentially impossible. Our results highlight how dynamical simulations of newly discovered exoplanetary systems could be used as an additional means to assess the potential targets of biosignature searches, and thereby help focus the search for life to the most promising targets.
\end{abstract}


%%%%%%%%%%%%%%%%%%%%%%%%%%%%%%%%%%%%%%%%%%%%%%%%%%%%%%%%%%%%%%%%%%%%

\section{Introduction}
\label{intro}

Through the coming decade, the next generation of astronomical observatories should yield a wealth of planets that, to a greater or lesser extent, seem to resemble the Earth \citep[e.g.][]{ric15,ary17,fra17}. At this point, there will be a significant investment of observational resources attempting to search for evidence of biosignatures on those alien worlds \citep[e.g.][]{Des02,Seg05,Kalt10,Rauer11,OMJ14}. But where should we look? By the time that we are capable of looking for the evidence of life on planets beyond the Solar system, we will likely have a vast catalogue of potential targets, from which the most promising must be chosen for that search \citep[e.g.][]{Turnbull03,Lammer09,HabRev,kop14,Barnes15,Cuntz16,Agnew17,Agnew18a,Agnew18b,Agnew19,Lingam18}.

The Exoplanet Era began in the latter stages of the last millennium, with the discovery of the first planets orbiting other stars \citep[e.g.][]{GammaCeph,Latham,psr1257,51peg}. The first planets discovered revealed that the diversity of planetary systems was far greater than we had previously imagined. Giant planets were found orbiting perilously close to their host stars - a population of planets that became known as `hot Jupiters' \citep[e.g.][]{51peg,Masset03,Bouchy05,HotJ2,HotJ1,albrecht12}. Planets were found orbiting pulsars \citep[e.g.][]{psr1257,psr1620,psr1719}, and others were found moving on highly elongated orbits, dramatically different to anything seen in the Solar system \citep[e.g.][]{Witt07,Tamuz08,Har15,HD76920}. 

As time has passed, every new generation of exoplanet discoveries has once again highlighted the great diversity of exoplanetary systems, driving home the concept that the planet formation process can yield an incredible variety of outcomes \citep[e.g.][]{ross458c,Candy,dense,Johns18}. This result has been strikingly driven home by the results from the {\textit {Kepler}} spacecraft, which carried out the first true exoplanetary census \citep[e.g.][]{Borucki10,Batalha13,Mullaly15}. Among its many other noteworthy discoveries, {\textit {Kepler}} revealed that `super-Earths' and `sub-Neptunes', classes of planet that are not represented in the Solar system \citep[e.g.][]{SE1,SE3,SE2,SE4,SE5}, are common. {\textit {Kepler}} also revealed that `dynamically packed' planetary systems are common - with the planets therein packed so tightly together that it would be impossible to have any other planets orbit between them \citep[e.g.][]{Pack1,Pack2,Trapp}.

At the same time, we have seen a revolution in our understanding of the formation and evolution of the Solar system. These developments include a new understanding of planet formation mechanisms \citep{adams2010}, the dynamical history and evolution of the orbits of Solar system bodies, \citep{duncan1993,tsiganis2005,nesvorny2018}, and the prevalence of water and water delivery mechanisms in the early Solar system \citep{encrenaz2008}. Such an understanding of our own planetary system is essential to placing exoplanetary systems in context, particularly given the diversity of orbital architectures that have been discovered in other planetary systems \citep{ford2014,kane2014,batygin2015,hatzes2016,raymond2018}. Many of the compact planetary system discoveries resulted from {\it Kepler} observations \citep{fang2012}, and it is expected that further insights into orbital architectures will result from the discoveries that will be made by the Transiting Exoplanet Survey Satellite ({\it TESS}). The parallel developments in both Solar and exoplanet system science are therefore gradually converging upon a consistent picture of planetary system architectures.

The next generation of exoplanet observatories will yield a vast fresh catch of new discoveries. Given the evidence that terrestrial planets are common in the cosmos \citep[e.g.][]{etaEarth,dress13}, it is likely that new discoveries, such as those from {\it TESS}, will include many planets in the super-Earth regime or smaller \citep[e.g.][]{ric15,sul15}. At the same time, new instruments such as the James Webb Space Telescope \citep{JWST, kem18} and the next generation of ultra-large ground based telescopes \citep[such as the European Extremely Large Telescope and the Giant Magellan Telescope;][]{EELT,GMT} are expected to be able to deliver the first measurements that could properly characterise such planets, and potentially detect any evidence of life upon them \citep[e.g.][]{beijwst,barhab,schjwst,proxjwst}. 

There is, however, a problem. Simply put, by the time we are ready to search for evidence of life on suspected `Earth-like' worlds, we will have far more potential targets to study than available resources to study them. For this reason, it is imperative to prepare the playing field, to examine the various facets that come together to make one planet more (or less) habitable than another \citep[e.g.][]{HabRev}. By doing so, we hope to inform the target selection for the search for life beyond the Solar system, helping researchers to target the most promising planets, and maximize the likelihood of a positive outcome.

In this light, a number of studies have begun to investigate the myriad factors that influence planetary habitability. Such studies range from studies of the impact of stellar variability and binarity on the climates of terrestrial planets with potentially temperate surface conditions \citep[e.g.][]{egg13,kan13,hag13,for14} to investigations of the role of giant planets in determining the impact rates on Habitable Zone (HZ) planets \citep[e.g.][]{FoF1,FoF2,FoF3,FoF4,LewisFoF,Grazier,Grazier2} to studies about the impact of planetary orbital and spin dynamics on the potential climate of a planet \citep{armst14,lins15,dei18b,georg18}. The planetary architecture plays a fundamental role in every one of those studies as gravitational forces induced mainly by planet-planet interactions alter the orbits of planets on astronomical timescales and thereby change the stellar radiation a planet receives --- one among many factors that could affect the suitability of a given planet as a host for life.

Throughout most of the Solar system's history, Earth's climate has remained within the range that allows liquid water to exist on the surface and thereby accommodated the development of life \citep{Mojzsis}. That is not to say that climatic conditions on our planet remained unchanged. The long-term climate of Earth is primarily driven by geological processes such as plate tectonics and volcanism that alter the composition of the atmosphere. On shorter timescales, periodic variations in climate (collectively known as Milankovitch cycles) are superimposed on the long-term trend. The gravitational interactions with objects in our Solar system induce the systematic flexing and tilting of Earth's orbit over time, and affect the planetary axial spin dynamics \citep[]{milank,berger94,lask04}. Subsequently, the distribution and to a lesser extent the total amount of solar flux received at the top of the atmosphere varies over timescales greater than 10$^4$ year, causing cyclic modifications to the global and regional environment \citep{berger78,berger94}. 

Perhaps the most famous example of Milankovitch cycles on Earth are the sequences of glacial and interglacial periods that have been particularly pronounced during the Late Pleistocene \citep[the past million years; e.g.][]{hays,imbrie,lis2005}. It should be noted that Earth's astronomical cycles have a relatively small amplitude, low frequency, and yet produce significant climate oscillations. Given that architectures of alien planetary systems are certain to be greatly different to that of the Solar system, some `exoEarths' are likely to experience astronomical forcing of greater variability \citep{dei18a,dei18b}. Such oscillations could have important implications for the suitability of a planet as a host for life, and may also impact the long-term survival of any existing life. Ample evidence exists that climate fluctuations on Earth have impacted the biodiversity, evolution, and migration of species \citep[e.g.][]{bennett,jansson,vandam}. More extreme climatic variations (such as Snowball Earth transitions) may even have facilitated the explosion of new life-forms during the Neoproterozoic \citep[e.g.][]{Hoffman,kirschvink}.

It is therefore clearly important to examine in detail the role that the architecture of planetary systems could have on the climates of potentially habitable worlds. Such studies could help to identify systems in which planets that would otherwise be considered eminently habitable could be ruled out as targets for the initial search for life beyond the Solar system on the basis of extreme climate variability on astronomical timescales, driven by interactions between the planets in that system.

In this work, we consider the influence of the architecture of our own planetary system on the astronomical cycles that Earth experiences. Essentially, we treat the Earth as though it were a candidate exoplanet, and examine the variability of its orbital elements on the basis of the architecture of its host planetary system. To sample potential architectures for the system, we choose a methodical approach, considering the question ``How would Earth's Milankovitch cycles change if Jupiter moved on a different orbit?". We expect that variations in Jupiter's orbit will be especially important in altering Earth's orbital cycles because the two leading terms in Earth's eccentricity solution are related to Jupiter's orbital geometry ~\citep{lask04}.  In Section~\ref{earth}, we offer a short refresher on the various Milankovitch cycles experienced by the Earth, before describing our methodology in Section~\ref{methods}. We present our results in Sections~\ref{jupiter} and 5, before discussing the implications of our results in Section~\ref{discussion}. Finally, we consider the implications of our work for the Rare Earth hypothesis in Section 7, before drawing our conclusions in Section~\ref{conclusions}, with a discussion of the direction in which we intend to take our future work.

%%%%%%%%%%%%%%%%%%%%%%%%%%%%%%%%%%%%%%%%%%%%%%%%%%%%%%%%%%%%%%%%%%%%

\section{Milankovitch Cycles of the Earth}
\label{earth}
The long-term ($>$10$^{4}$ year) variability of Earth's climate is driven by intricate interactions between our planet's orbital evolution, spin dynamics, the spatial distribution of continents, oceans, variations in biogeochemistry, the occurrence of ice sheets, and many other factors \citep[see e.g.][]{Cronin}. To fully model such long-term climate behaviour for newly discovered exoplanets is currently beyond us, as most of the required information will not be available. To put this complexity in context, we describe the current understanding of astronomically-forced climate cycles experienced by the Earth.

Three astronomical cycles are typically discussed in the context of Earth's climate, namely eccentricity (\textit{e}, the ellipticity of Earth's orbit), obliquity ($\epsilon$, the angle of planetary axial tilt relative to the orbital plane) and climatic precession (\textit{e} sin$\varpi$, where $\varpi$ is the longitude of perihelion{\footnote{$\varpi$ is the sum of the longitude of the ascending node ($\Omega$) and the argument of perihelion ($\omega$). The longitude of the ascending node of orbits in the Solar system is measured from a specific reference direction, known as the first point of Aries, which marks the location of the Sun in the sky at the time of the Vernal Equinox. The direction of this reference varies with time as the Earth's axis precesses, which astronomers address by providing orbital elements that are accurate at a specific reference epoch. The argument of perihelion is then measured from the ascending node to the perihelion in the orbital plane.}}). 

Climatic precession describes which hemisphere faces toward the Sun during a particular season and thus controls the spatial distribution of incident solar flux that is responsible for the seasonal contrast between hemispheres. When the Northern Hemisphere (NH) faces toward the Sun at perihelion, NH summers are particularly warm, while winters are extremely cool. At the same time, the Southern Hemisphere experiences reduced seasonal disparity. A precession cycle is approximately 23 kiloyear (kyr) long, meaning that the hemisphere experiencing maximal seasonal contrast alternates every $\sim$11 kyr. Eccentricity and precession are tightly connected in the sense that a perfectly circular orbit (\textit{e} = 0) results in an equal amount of insolation received by both hemispheres, which minimizes interhemispheric differences. Even though the eccentricity of Earth may reach values close to zero every $\sim$100 kyr, hemispheric contrast never completely disappears. The oceans and continents have a different albedo, heat capacity, and thermal inertia; the asymmetric continental distribution therefore maintains the hemispheric contrast to some extent, even at times of near-zero orbital eccentricity.

The $\sim$100 kyr (short) and $\sim$400 kyr (long) eccentricity cycles not only modulate climatic precession effects, they also slightly alter the total amount of insolation Earth receives. The annual mean solar flux (F) at the top of the atmosphere scales with orbital eccentricity such that F $\propto$ (1-\textit{e}$^{2})^{-0.5}$. Earth's eccentricity varies between $\sim$ 0 and 0.06, which is equivalent to a difference in radiative forcing of $\sim$0.5 W~m$^{-2}$ on astronomical timescales \citep{lask04,berger78,berger94}.

Despite these small variations in eccentricity, power spectra of paleoclimate data often reflect a strong imprint of eccentricity frequencies. The glacial-interglacial cycles of the late Pleistocene appear to be eccentricity-paced, whilst $\sim$100 and $\sim$400 kyr cycles are also prominently present in mid- and early Cenozoic and Mesozoic geological records \citep[e.g.][]{herb86,lourens05,KT2014}. The discrepancy between the small changes in annual mean insolation and the relatively large climate consequences that result could imply that processes intrinsic to Earth may lead to a highly nonlinear response between insolation and climate \citep{hays,clem}. Positive feedbacks, for example such as the ice-albedo feedback \citep{curry,mcgehee}, may drive the climate state of a planet into its extremes, while negative feedbacks, such as CO$_2$ consumption through weathering \citep{walker} can dampen the effects of cyclic insolation variation on timescales greater than 10$^4$ year. 

We should note that there are still many unknowns in the theory of Milankovitch forcing. One of the main ideas is that the $\sim$100 kyr glacial cycles of the late Pleistocene are not directly related to eccentricity, but rather indirectly through the modulation of precession effects \citep{raymo97,ridgwell,maslin}. The long ($\sim10^5$ yr) residence time of carbon in the oceans and atmosphere may also transfer power from precession frequencies to the longer eccentricity frequencies \citep{zeebe}. Alternative explanations for the $\sim$100 kyr cycles exist that invoke intrinsic feedback mechanisms that could explain a natural climate oscillation on $\sim$100 kyr timescales due to inertia in parts of the Earth system that are completely unrelated to astronomical forcing \citep{saltz,wunsch2003,nie08}. At the same time, it has been suggested that obliquity forcing, rather than eccentricity variation, is the main driver for climate oscillations \citep[e.g.][]{huybers05,raymo06}, on the basis that obliquity has a more direct impact on the mean insolation at high latitudes than does eccentricity. It should be noted, however, that variations in eccentricity would serve to modulate the impact of such obliquity variations.

The Earth's obliquity, or axial tilt, oscillates between 21.5$^\circ$ and 24.5$^\circ$ with a period of $\sim$~41 kyr. This expands and shrinks the polar circles, alters high-latitude insolation patterns and subsequently leads to the waxing and waning of ice caps. 41 kyr cycles in paleoclimate records are particularly pronounced in the Pliocene and early Pleistocene \citep{gildor,naish,lourens}.

Most likely, all of the above suggestions play a role in the periodic climate oscillations on Earth, but the relative importance of those element shifts depending on the state of the planet during a given period in geological history. For instance, the presence and distribution of ice sheets \citep{raymo06}, or the background climate state \citep{berger1999} could shift the spectral signal of the oscillations.

%To illustrate this, one can consider the behaviour of  Obliquity variations played a particularly important role during the early Pleistocene and Pliocene periods, when Earth's climate cycles were dominated by obliquity frequencies .
%The combined effect of changes in the orientation of Earth's spin axis (axial precession) and the rotation of Earth's perihelion relative to the background stars (apsidal precession) is expressed as climatic precession oscillating with a period of approximately 23 kyr. 
%In case of a perfectly circular orbit, the seasonal contrast between hemispheres is minimised. This dependency of the precession effect on eccentricity is integrated in the precession index as \textit{e} sin$\varpi$, where sin$\varpi$ denotes the longitude of perihelion. 

It is clear that astronomical forcing contributed to the oscillations in Earth's climate not only in the more recent time when polar ice caps could amplify the effects of insolation changes, but also in the more distant geological history when the Earth surface was mostly devoid of ice --- despite the minor amplitudes of the eccentricity and obliquity cycles of Earth \citep[e.g.][]{zachos}.

Given the expected wide diversity in exoplanetary system architectures \citep{win15}, it is reasonable to assume that some planets orbiting other stars undergo more extreme orbital variations and therefore experience dramatic spatiotemporal insolation changes, potentially having important implications for the climatic environment of an exoplanet and its habitability. Since the obliquity and precession dynamics of terrestrial exoplanets can not currently be observationally constrained, we focus here on the orbital parameters that are measurable, namely the planet's orbital eccentricity and inclination (the tilt of the orbit relative to the reference plane). We would like to stress, however, that the orbital and spin dynamics are intimately connected. Any changes in the orbital inclination or eccentricity due to planet-planet interactions will directly translate, although non-linearly, into perturbations to the spin dynamics of that planet \citep[e.g.][] {kin75,lask04}.

The calculations to estimate the evolution of the planetary spin dynamics are extremely complex and output is highly sensitive to the multitude of input variables. Not only is the angular momentum altered by variations in the orbital motion, it is also affected by tidal forces \citep{lask93} and internal planetary processes such as core-mantle friction \citep{neron}, atmospheric tides \citep{barnes83,voll96}, mass displacement from plate tectonics or mantle convection \citep{ward79,forte} and climatic friction \citep{dehant90,rub90,ito95}. Such information will likely never be obtained for exoplanets.

For that reason, we explore the manner in which the orbital evolution of Earth responds to architectural changes to the Solar system. Specifically, in this work we consider the impact of the orbit of the Solar system's most massive planet, Jupiter, on the Earth's orbital cycles. 

%In dynamical terms, the three Milankovitch cycles relate to five modes of motion that can be split into two groups. (1) The orbital group, including eccentricity, orbital inclination and apsidal precession, is affected by planet-planet interactions. (2) The spin group, tied to axial spin momentum, consists of obliquity and axial precession \citep[e.g.][]{lask93,wil94,lask04,walthamaxis}. From an exoplanet perspective, only the orbital parameters can be measured or inferred from observed data, but 

%,  -- then later, can say something like 'When we want to learn about the climate on newly discovered exoplanets, the amount of information we have will be limited to their orbital characteristics. In this work, we examine how .... also something like 'On Earth, the orbital variations are just part of the story, and interact non-linearly with the other properties... then talk about the feedbacks etc.

%%%%%%%%%%%%%%%%%%%%%%%%%%%%%%%%%%%%%%%%%%%%%%%%%%%%%%%%%%%%%%%%%%%%

\section{Dynamical Simulation and Methodology}
\label{methods}

To examine how Earth's orbital Milankovitch cycles would be altered were the Solar system's architecture markedly different, we carried out an extensive suite of $n$-body simulations, using the Hybrid integrator within the dynamical integration package {\sc Mercury} \citep{Mercury}. Each individual simulation followed the dynamical evolution of the eight Solar system planets for a period of ten million years, with the instantaneous orbital elements of the planets written to file every thousand years. An integration time-step of one day was used to ensure that the integrations were as accurate as possible.

In addition, {\sc Mercury} was modified by the addition of a user-defined force in order to take account of first-order post-Newtonian relativistic corrections \citep{Gil08}. This correction allows the orbital behaviour of the innermost planets to modeled accurately, ensuring that our results fairly reflect the physical reality of the orbital evolution of the planets in question. 

A total of 159,201 individual simulations were carried out, in which the initial orbits of the planets Mercury, Venus, Earth, Mars, Saturn, Uranus and Neptune were the same. The only elements that were changed from one simulation to the next were the initial semi-major axis, $a$, and eccentricity, $e$, of Jupiter's orbit. The process by which the simulations were created followed that established for studies of the stability of exoplanetary systems and Solar system small bodies \citep[e.g.][]{QR322,Anchises,24Sex,30177}, such that the semi-major axis and eccentricity of Jupiter were each varied in regular steps, creating a rectangular grid of solutions in $a$--$e$ space for Jupiter's orbit.

The orbital elements for the planets were obtained from the Horizons DE431 ephemeris, and converted from Cartesian to Keplerian coordinates through {\sc Mercury}, by running a short simulation and outputting the Keplerian elements at $t = 0$. These Keplerian elements were then used as the basis for our grid of Jovian orbital solutions. In total, we tested 399 unique values of Jupiter's semi-major axis, covering a 4.0~au range centred on the initial semi-major axis of Jupiter in `our' Solar system ($a \sim 5.203$~au). For each of these 399 semi-major axes, we tested 399 unique values of Jovian orbital eccentricity, ranging from circular orbits ($e = 0.0$) to ones with moderate eccentricity ($e = 0.4$). Whilst such a high upper bound might seem unusually high, we note that there is a growing body of exoplanets that have been found on orbits far more eccentric than this \citep[e.g.][]{Witt07,Tamuz08,Har15,HD76920}. Even in the Solar system, it has been suggested that the eccentricity of Mercury's orbit can exceed this value, as part of its own long-term periodic variability \citep[e.g.][]{Strom,Correia}. 

Individual simulations were halted early if any of the planets was so perturbed that it collided with the Sun, another planet, or reached a heliocentric distance of 40~au. We flagged those simulations that contained architectures that proved dynamically unfeasible, and recorded the time within the integration at which the simulations were halted. Once the remaining (stable) simulations were complete, we extracted the evolution of Earth's orbital elements at 1,000 year intervals to determine the frequency and amplitude of the variations that occurred in Earth's orbit. 

\begin{figure*}
\centering
  \includegraphics[width=0.9\textwidth]{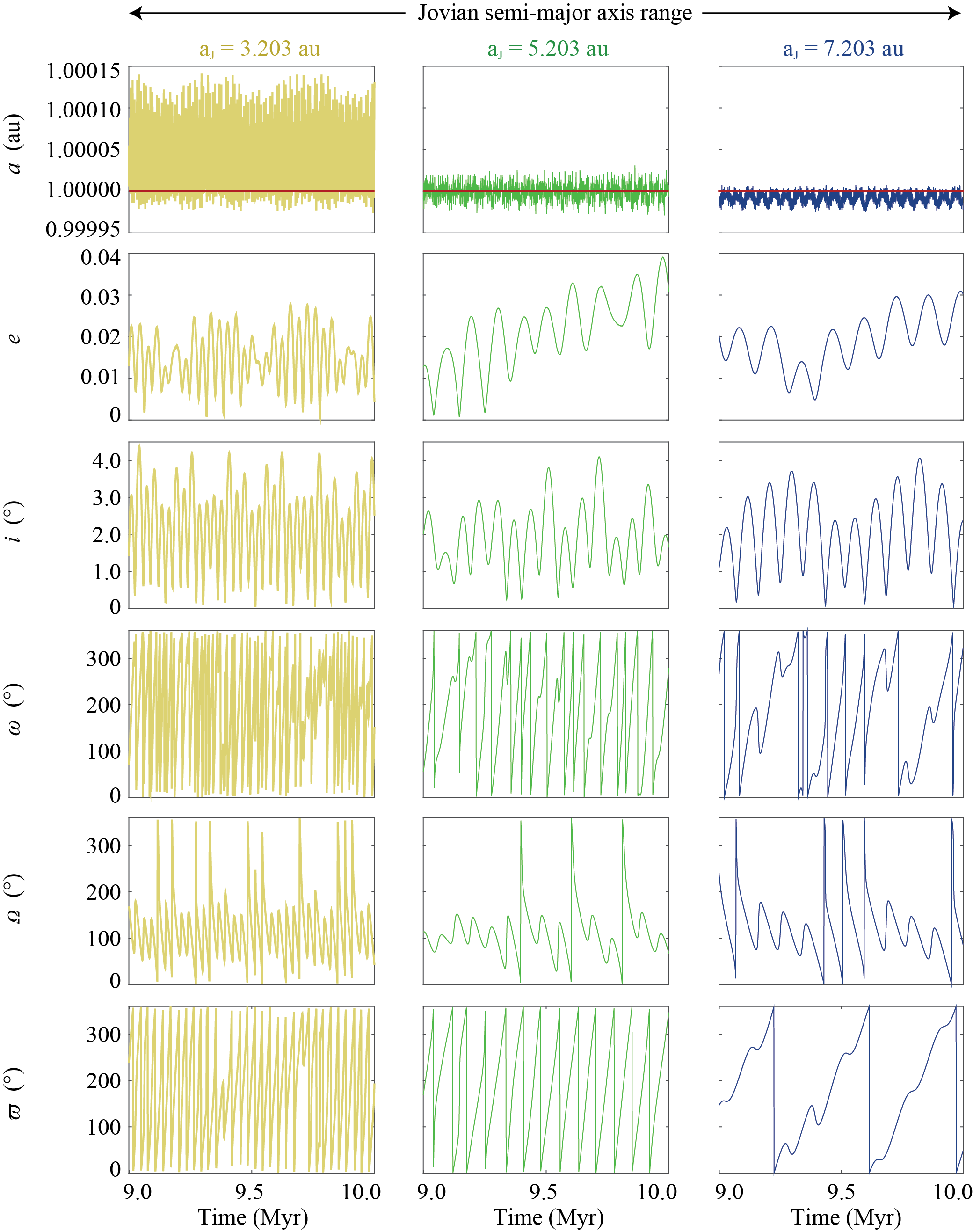} \\ 
\caption{Plot showing the variations in the Earth's orbital elements for a period that spans the last million years of the ten million year simulations for three exemplar scenarios. The first three rows show (top to bottom) the evolution of Earth's semi-major axis, $a$, eccentricity, $e$, and its orbital inclination, $i$. The fourth and fifth rows the argument of the Earth's perihelion, $\omega$ and the longitude of the ascending node of Earth's orbit, $\Omega$. The final row shows the longitude of Earth's perihelion, $\bar{\omega} = \omega + \Omega$. The scenarios shown highlight the impact of moving Jupiter's initial orbit to a new semi-major axis. The left column shows data obtained with Jupiter moved inwards by 2 au (i.e. $a_J = 3.203$~au), the central column shows data with Jupiter at its current semi-major axis (i.e. $a_J = 5.203$~au), and the right column shows data for Jupiter moved outwards by 2 au (i.e. $a_J = 7.203$ au), In all cases, the initial orbital eccentricity of Jupiter was set to zero, and the orbital elements of all other planets were set to their modern, canonical values.}
\label{exemplar}
\end{figure*}

In Figure~\ref{exemplar}, we show the evolution of the Earth's orbital elements over the last million year of our simulations for three exemplar cases. The three scenarios feature `Jupiters' initially located at 3.203~au (left), 5.203~au (centre) and 7.203~au (right), while Jupiter's initial orbital eccentricity was set to 0.0 and initial values of the four other orbital elements were set to their canonical values. This figure highlights the degree to which changes in Jupiter's orbit can impact both the amplitude and frequency of the Earth's orbital cycles.

\begin{figure}
\begin{tabular}{c}
 \includegraphics[width=0.49\textwidth]{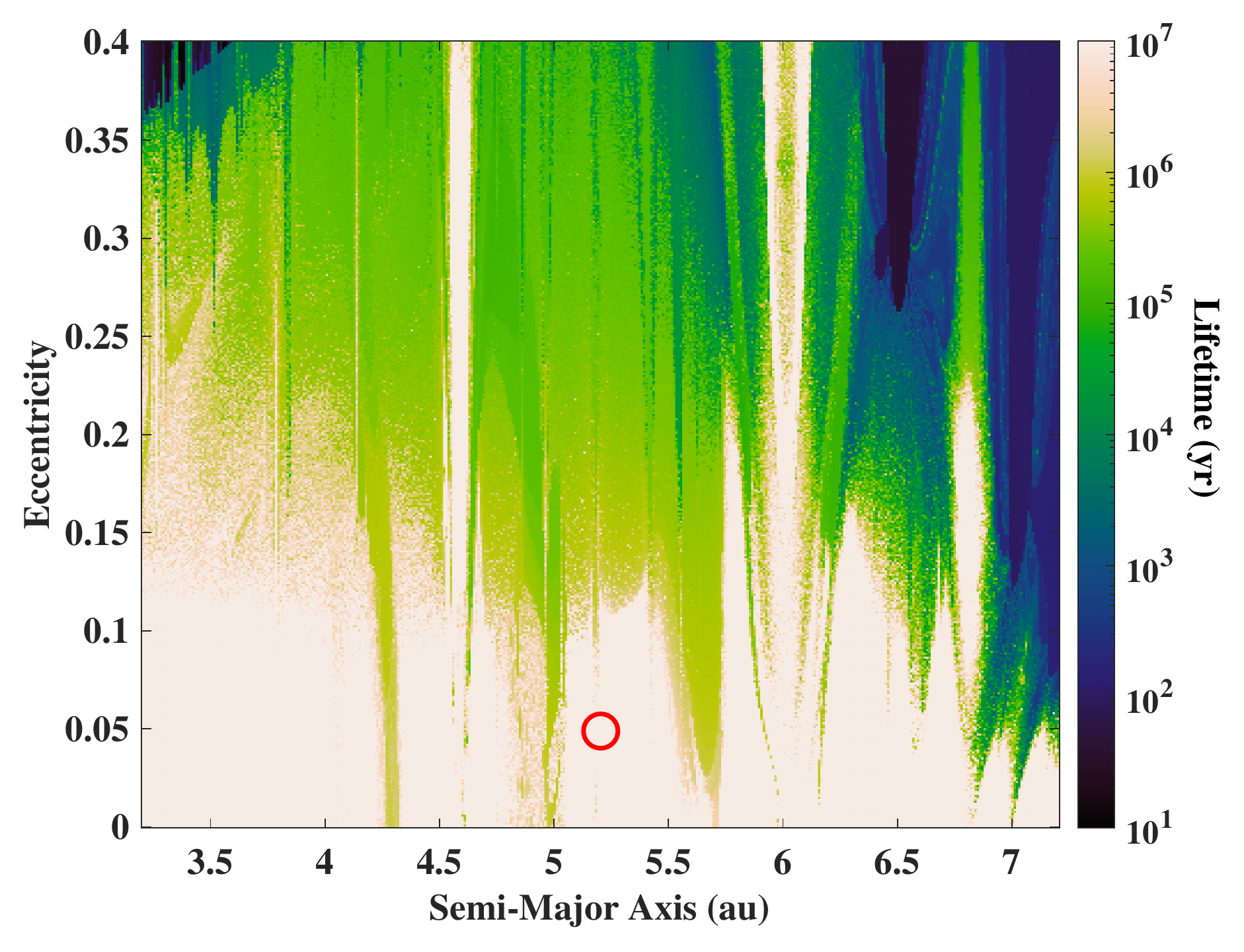} \\
\end{tabular}
\caption{The dynamical stability of the Solar system over the ten million years of our simulations, as a function of the initial semi-major axis, $a$, and eccentricity, $e$, of Jupiter's orbit. The open red circle marks the location of Jupiter in the modern Solar system, for reference. Of 159,201 simulations carried out, just 41,652 survived until the end of our integrations, with the remainder ($\sim$74\% of the tested sample) becoming unstable before the integrations were complete.}
\label{stability}
\end{figure}

Of 159,201 unique realisations of the Solar system simulated in this work, the majority ($\sim 74\%$) proved unstable. The stability of the Solar system in our simulations is shown in Figure~\ref{stability}. It is clear, particularly when the data are plotted on a linear scale, that the stability of the Solar system is a strong function of Jupiter's orbital eccentricity, but even at low eccentricities, there are regions where no stable solutions were found. Equally, two narrow regions can be seen where stable solutions exist across the full range of eccentricity tested in this work -- the result of the stabilising influence of resonant interactions between Jupiter and Saturn. 

From Figure~\ref{stability}, it is apparent that instability occurs on a variety of timescales - and the further you move from the 'stable' solutions in our work, the more rapidly things become unstable. As a result, it seems highly likely that scenarios on the fringe of stability, which in our simulations proved stable on 10 Myr timescales, would likely become unstable on timescales of tens, hundreds, or thousands of millions of years. Whilst such instability would be of great interest, the computational challenges involved in integrating our data for one or two orders of magnitude longer make a detailed study of such edge cases impractical. Nonetheless, we caution readers that the true extent of the 'unstable' region is likely slightly larger than that seen in our data, as a result of the dynamical timescale over which our integrations were performed.

\begin{figure*}
\begin{tabular}{cc}
  \includegraphics[width=0.5\textwidth]{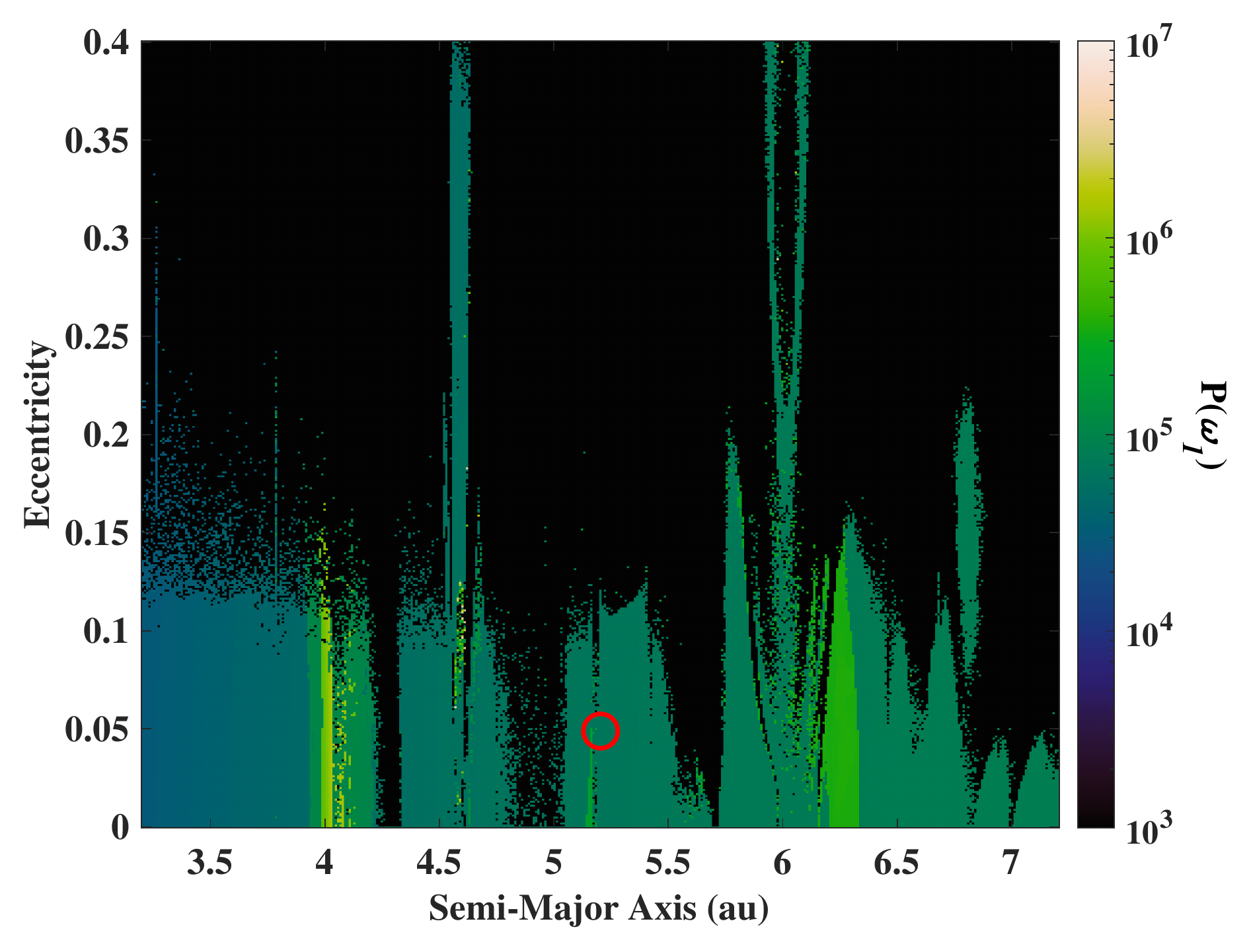} &  \includegraphics[width=0.5 \textwidth]{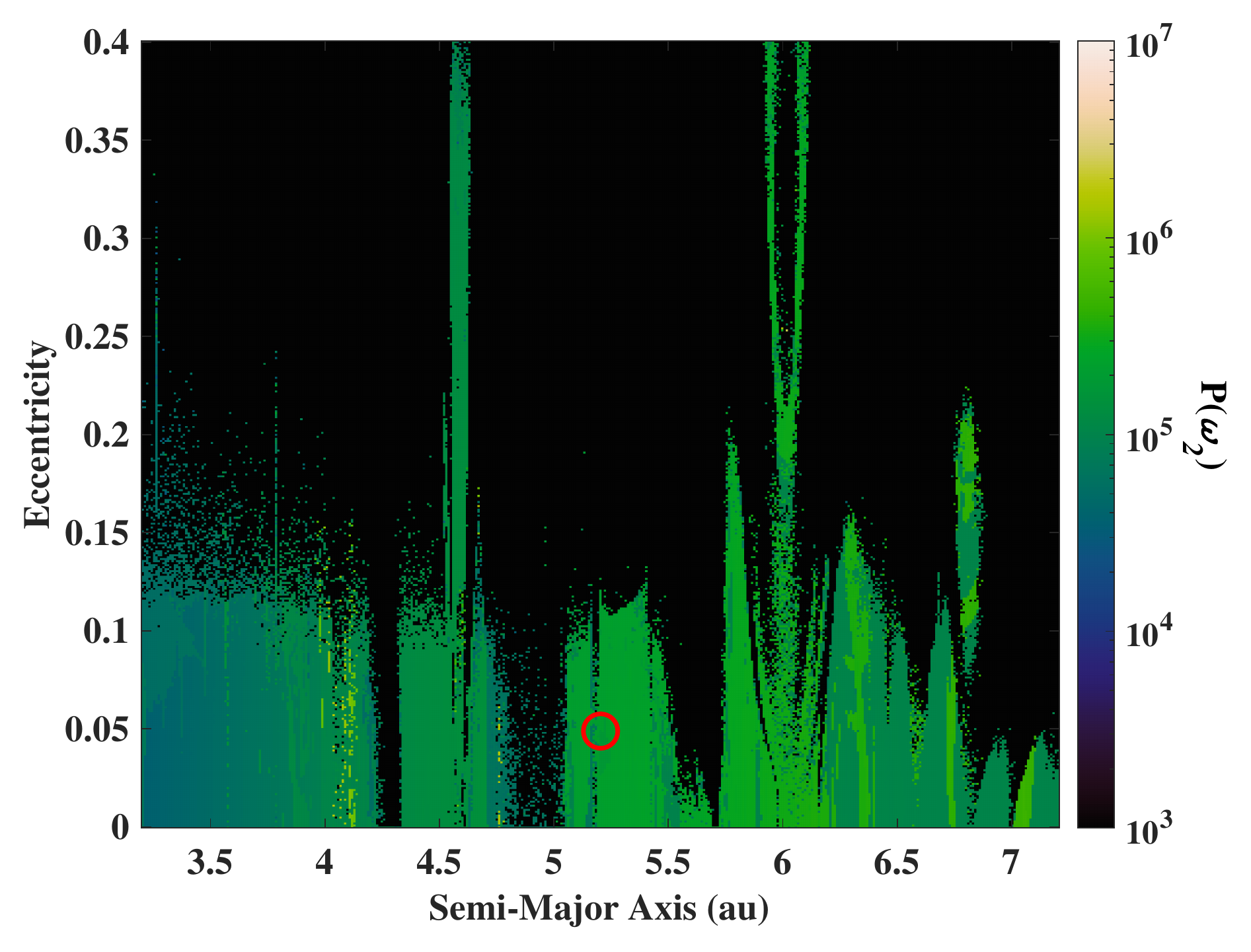} \\
\end{tabular}
\caption{The period of the two dominant oscillations in the argument of Earth's perihelion, as a function of Jupiter's initial semi-major axis, $a$, and eccentricity, $e$. The left-hand plot shows the period (in years) of the most dominant frequency, while the right-hand plot shows the period (in years) of the second-most dominant frequency.}
\label{omega}
\end{figure*}

For the rest of our analysis, we ignore the unstable regions of the plot, and solely focus on those regions where the system remains dynamically stable. For each of those stable solutions, we can determine the degree to which the Earth's orbital parameters vary with time - in both their amplitude and frequency. The results are visualized in Figures~\ref{powerspectrum} -- \ref{omega}.

%
% Replot Linear plot; modify caption and specify in the text the MMR protection
%

%%%%%%%%%%%%%%%%%%%%%%%%%%%%%%%%%%%%%%%%%%%%%%%%%%%%%%%%%%%%%%%%%%%%

\section{Spectral Analysis of The Dynamical Impact of Jupiter}
\label{jupiter}

\begin{figure*}
\begin{center}
\begin{tabular}{c}
\includegraphics[width=1.0\textwidth]{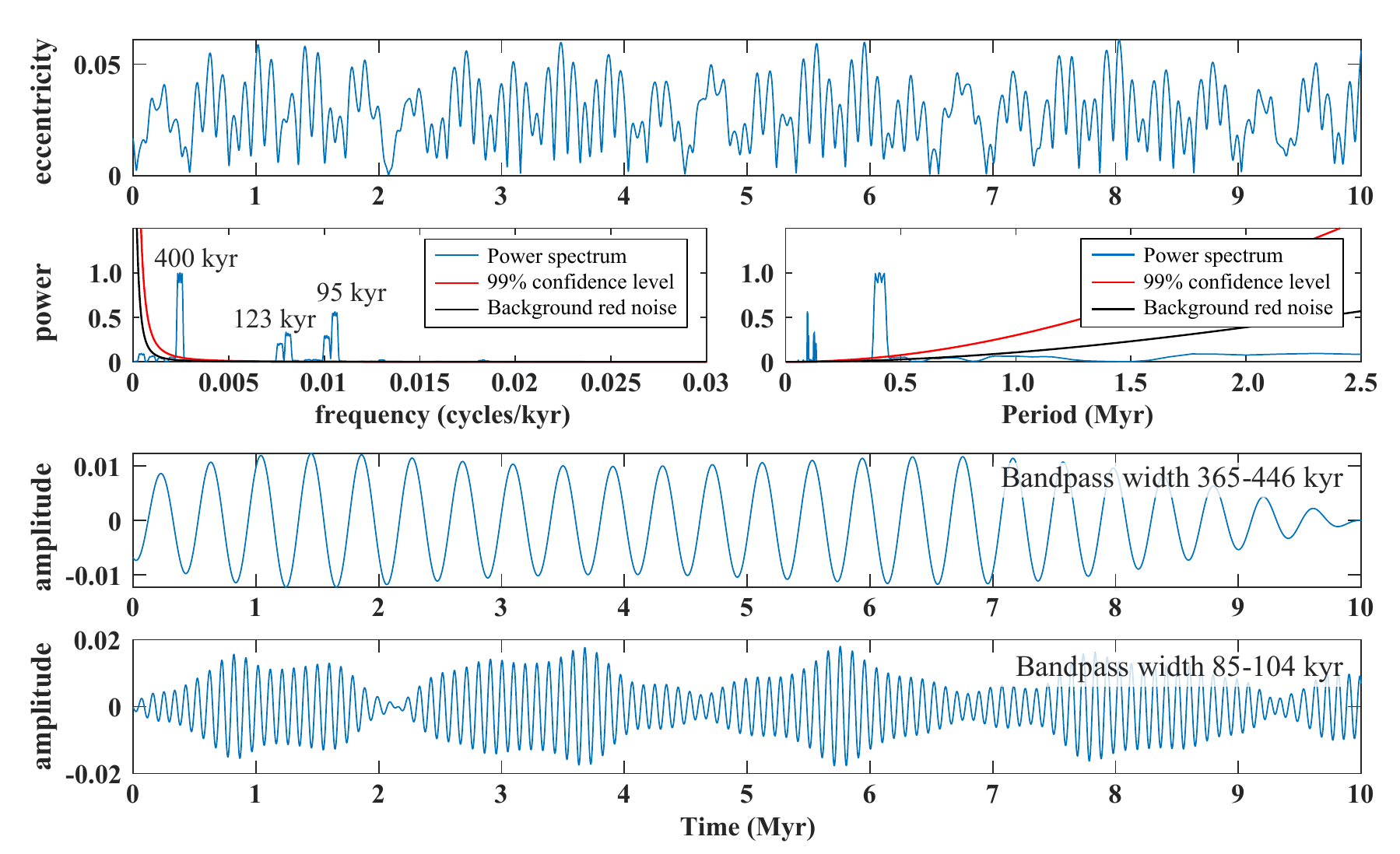} \\
\end{tabular}
\end{center}
\caption{Illustration of the means by which the amplitude and frequency of the dominant periods of Earth's orbital evolution are determined in this work. The top panel shows the time variation of Earth's orbital eccentricity for the individual member of our 41,652 dynamically stable simulations that most closely resembles the current configuration of the Solar system, with Jupiter located at 5.203 au and eccentricity of 0.049. The two panels in the second row show the normalised multitaper power spectrum of the time series (blue), in the frequency domain (left) and the period domain (right). The black line indicates the background robust AR1 red noise model, and the red line shows the 99\% confidence level above that red noise. The three most powerful and significant cycles detected have periods of 400, 123 and 95 kyr, consistent with the eccentricity oscillations that we experience on Earth. The plot on the third and fourth row show the bandpass filtered signal for the 400 and 95 kyr cycle, calculated with a bandpass width that is 10\% of the period.}
\label{powerspectrum}
\end{figure*}

\begin{figure*}
	\begin{center}
  		\includegraphics[width=1.0\textwidth]{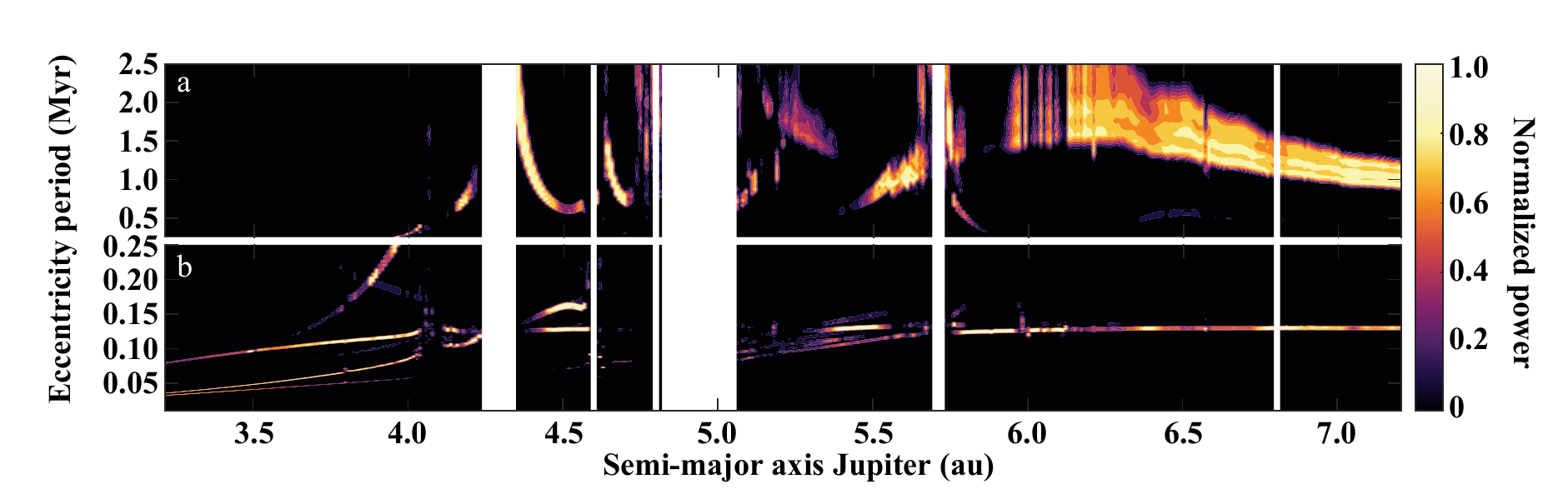} \\
  		\includegraphics[width=1.0\textwidth]{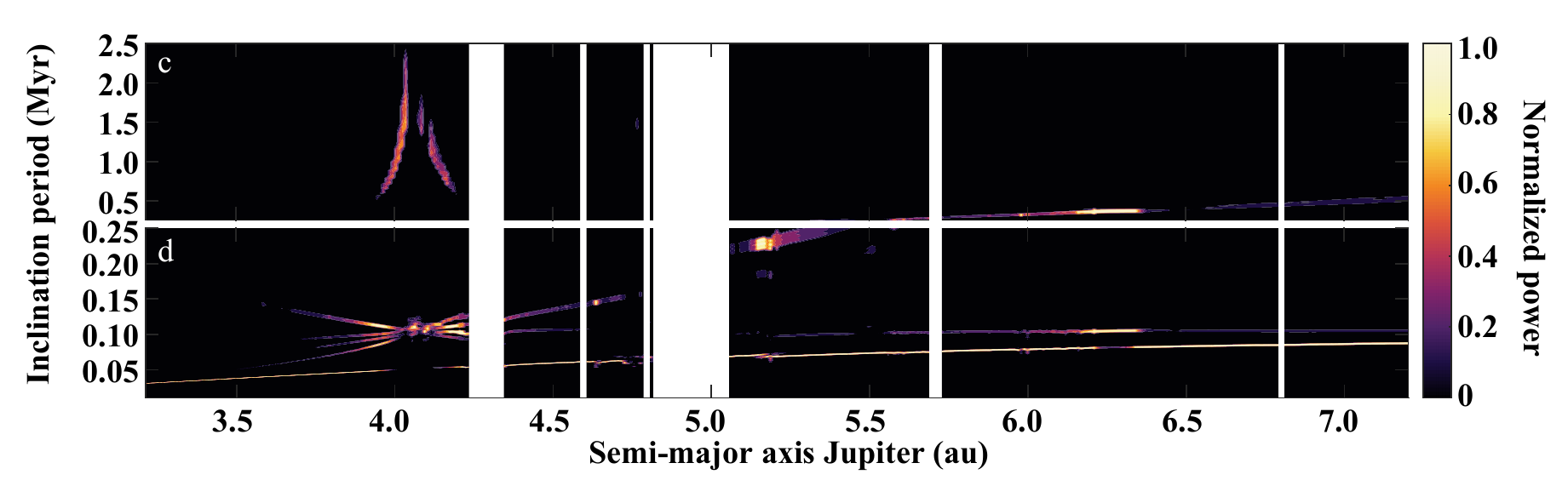}
	\end{center}
	\caption{Spectral analysis of Earth's eccentricity and orbital inclination as a function of Jupiter's semi-major axis, for 399 simulations in which Jupiter's eccentricity ($e_J$) is fixed at 0.0. Simulations where $e_J \neq$ 0.0, have similar cycles. For all 399 outputs, normalized multitaper power spectra are calculated and color-coded only when frequencies exceed the 99$\%$ confidence level. Dashed areas indicate unstable regions. A) Significant spectral power for Earth's eccentricity cycles ranging from 2.5 Myr to 250 kyr. B) Significant spectral power for Earth's eccentricity cycles $<$250 kyr. C) Significant spectral power for Earth's cycles in orbital inclination ranging from 2.5 Myr to 250 kyr. D) Significant spectral power for Earth's cycles in orbital inclination $<$250 kyr. *Note the changing y-axis between panels A and B, C and D.}
	\label{fourierfigs}
\end{figure*}

Spectral analysis was performed on each of the stable dynamical simulations to extract the primary and secondary periods and amplitudes for Earth's orbital elements. The $R$ package `Astrochron` was used to perform a multitaper method spectral analysis of our data using a time-bandwidth product of 2.0 (Meyers, 2014). To confirm the significance of spectral peaks, autoregressive (AR1) red noise models were generated for each individual time series. Spectral peaks were considered to be statistically significant when they rise above the 99$\%$ confidence level (Figure~\ref{powerspectrum} and \ref{fourierfigs}).
The maximum amplitudes of the two most pronounced and statistically significant cycles were estimated by applying a bandpass filter to the original time series data. The width of the filter is 10\% of the period that is associated with the cycle. For instance, a cycle with a period of 100 kyr will have a bandpass width of 10 kyr, covering 95 to 105 kyr periods. All significant cycles that have periods within this range are considered to have a similar origin. In this manner, we accounted for the temporal variability of a cycle over the simulated 10 million years. 

For validation, we applied these methods to the simulation in which Jupiter's location and eccentricity most closely resemble its 'real' values (Figure~\ref{powerspectrum}). The most pronounced cycles that result for Earth's eccentricity have periods of 400, 123 and 94 kyr, consistent with earlier calculations \citep[e.g.][]{lask04}. Likewise, the associated simulated amplitudes for Earth's eccentricity are also in good agreement with the expected values.

Spectral analysis reveals that the periods in Earth's orbital cycles are mainly affected by the semi-major axis of Jupiter $a_J$ (as can be seen in Figures~\ref{fourierfigs}, \ref{ecc_2dw}, and \ref{inc_2} - moving from left to right within a subplot). The influence of Jupiter's eccentricity $e_J$ is minor (e.g. Figures~\ref{ecc_2dw} and \ref{inc_2} - moving from bottom to top within a subplot). To examine this in more detail, we focus on the variability in Earth's orbital parameters over a range of $a_J$, while keeping $e_J$ constant at zero (Figure~\ref{fourierfigs}).

When Jupiter is initially placed close to the Sun at $a_J<4$ au, Earth's eccentricity cycles predominantly have short periods of 50-100 kyr. In scenarios where Jupiter is more distant, the dominant eccentricity cycles tend to have periods of 100-150 kyr and 1-2 Myr, as can be seen in panels A and B of Figure~\ref{fourierfigs} and in Figure~\ref{ecc_2dw}. These cycles appear to be disrupted when Jupiter is moved to 4.1 au, but this does not result in system instability within the simulated 10 million years. However, the cycles in Earth's eccentricity at this location exhibit aperiodic and relatively large-amplitude variations that suggest the system may become truly dynamically unstable after the 10 million years of our integrations. 

Rapid changes in the long-period oscillations in orbital eccentricity occur around 4.25, 4.6, 4.8, 5.05 and 5.7~au, demonstrating that the periods of the longest eccentricity cycles are sensitive to minor changes in Jupiter's semi-major axis just before and after an unstable region. This does not apply for the shorter cycles that are relatively stable throughout. For $a_J$ $\sim$ 5.2 to 7.2~au, the periodicity of the short eccentricity cycles remains at $\sim$140 kyr. The long eccentricity cycles vary from 500 kyr to 2.5 Myr, depending on Jupiter's location. 

Cycles in Earth's orbital inclination are generally shorter than eccentricity cycles (Figure~\ref{fourierfigs}, panels C and D). The period of the dominant cycle increases from 30 to 85 kyr when Jupiter is moved outward. The disruption around 4.1~au is clearly visible. Multiple short cycles converge at this point, while a strong peak appears in the longest cycle that rapidly rises from 500 kyr to 2.5 Myr, before falling back to $\sim$500 kyr. At this location, the Earth's inclination cycles exhibit large oscillations (sometimes greater than 10$^{\circ}$). However, in contrast to the aperiodic variability in Earth's eccentricity, these cycles appear to be steady over time throughout the 10 million year simulation. 

Aside from the most dominant 30-85 kyr cycle, two other orbital inclination cycles occur that arise in the vicinity of the 4.1~au inconsistency. We observe a 100 kyr cycle that remains approximately constant, and a cycle that starts with a period of $\sim$100 kyr and gradually increases to 500 kyr as the Jupiter-Sun distance increases to 7.2 au. 

Multiple significant cycles coexist in the variability of Earth's orbital eccentricity and inclination at any given Jupiter-Sun distance. This is because Jupiter is not the only planet that affects Earth's orbit - the other planets in the Solar system also interact gravitationally with the Earth, as well as with Jupiter. In scenarios where Jupiter is initially placed close to the Sun, the periods of Earth's orbital eccentricity and inclination cycles are short and remarkably similar. Two orbital parameters with similar periods can potentially lead to interesting climatic behavior as external forcing and subsequent climate feedbacks may either cancel out or reinforce each other, depending on the phase of the variation of the parameters. As the parameters move in and out of phase, their combined effects could 'beat', causing significantly larger variability than would otherwise be expected. 

%%%%%%%%%%%%%%%%%%%%%%%%%%%%%%%%%%%%%%%%%%%%%%%%%%%%%%%%%%%%%%%%%%%%

\section{Orbital Characteristics}
\label{OrbitalCharacteristics}

The degree to which Jupiter influences Earth's orbital cycles is clearly apparent upon examination of the three exemplar simulations shown in Figure~\ref{exemplar}. Both the amplitude and frequency of the cyclic variations in Earth's orbital elements are affected by Jupiter's orbit, which would change the spatiotemporal distribution of incident Solar flux and thereby alter the seasonal patterns experienced by Earth. 

\begin{figure*}
\begin{tabular}{cc}
  \includegraphics[width=0.5\textwidth]{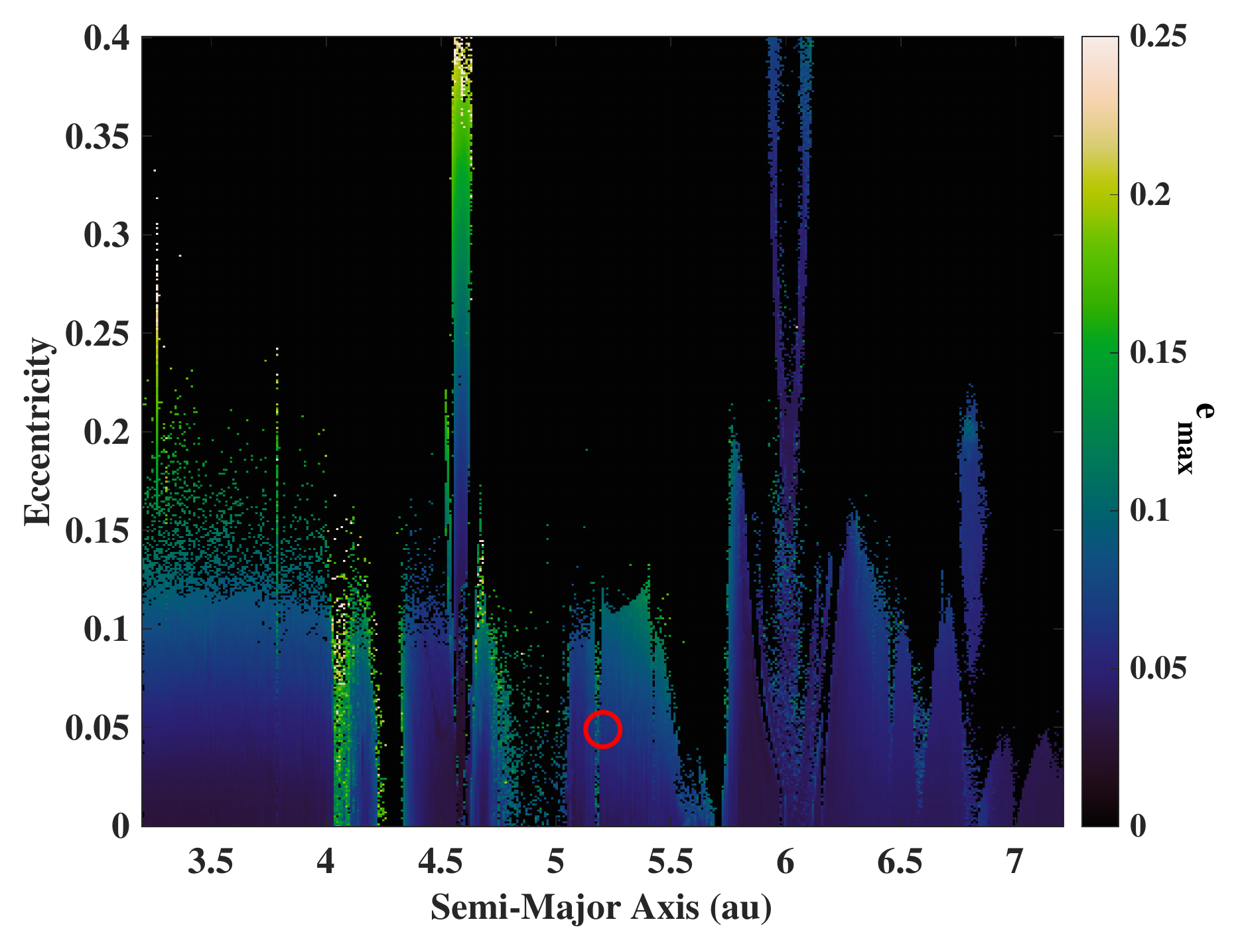} & \includegraphics[width=0.5\textwidth]{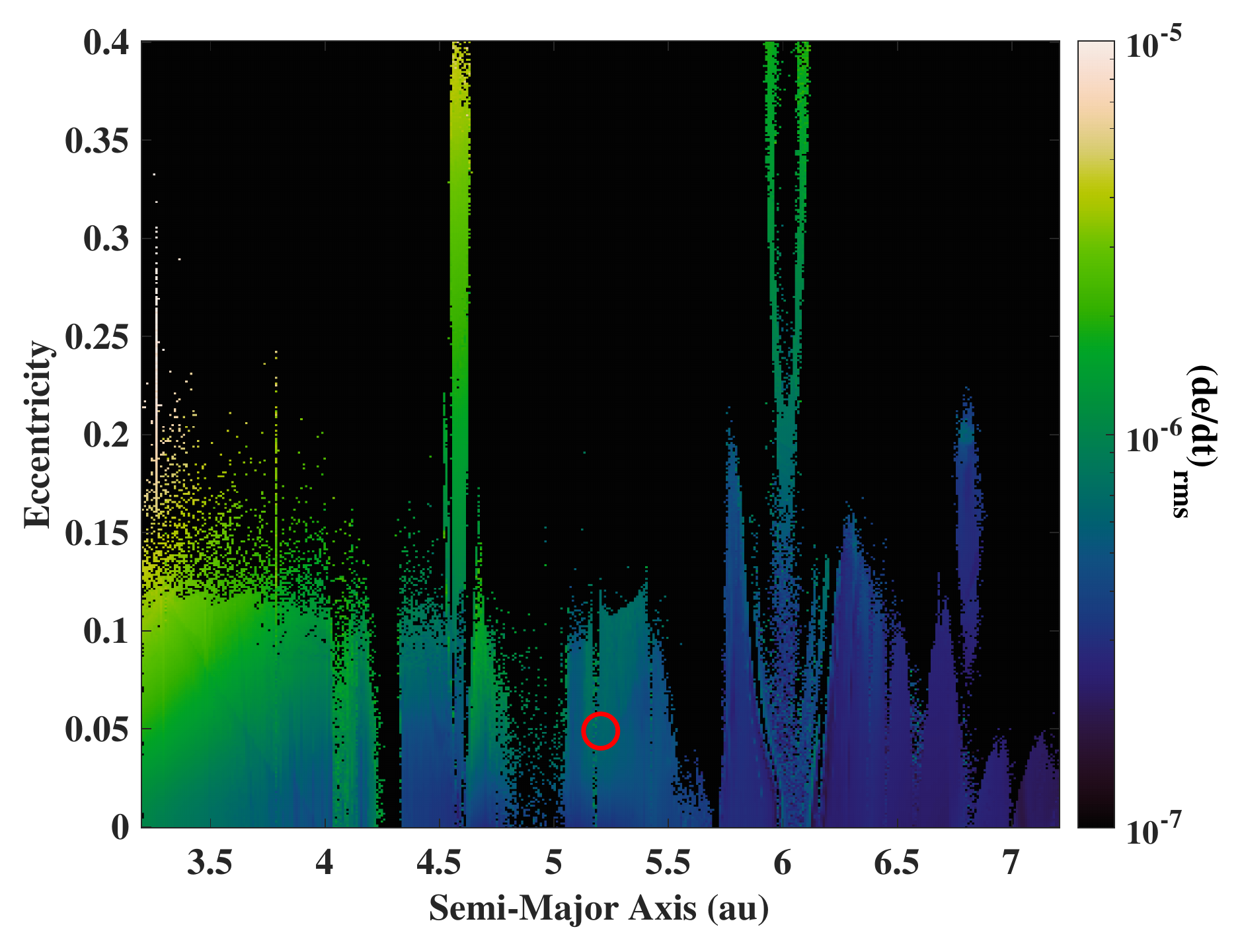}\\  \includegraphics[width=0.5 \textwidth]{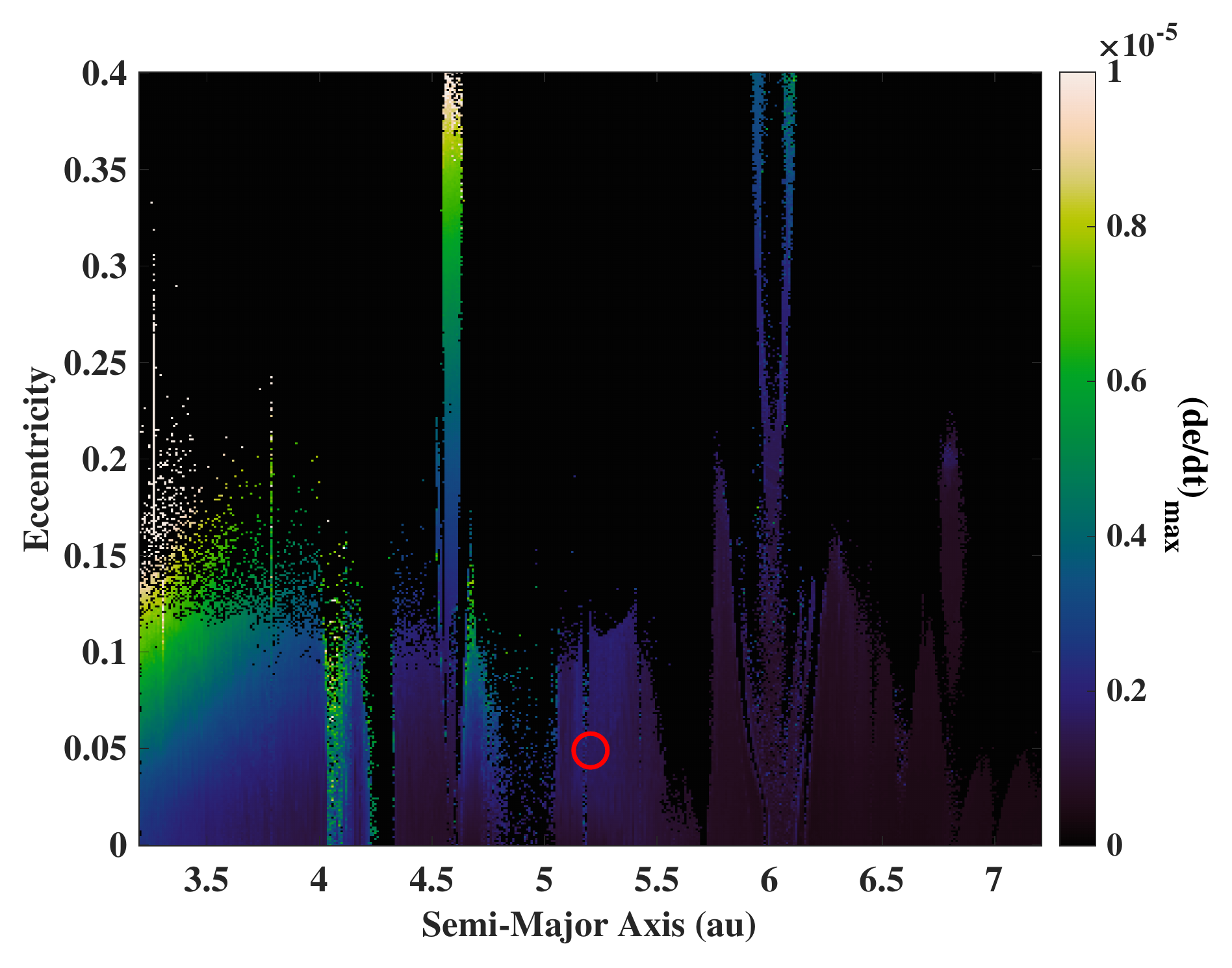} & \includegraphics[width=0.5 \textwidth]{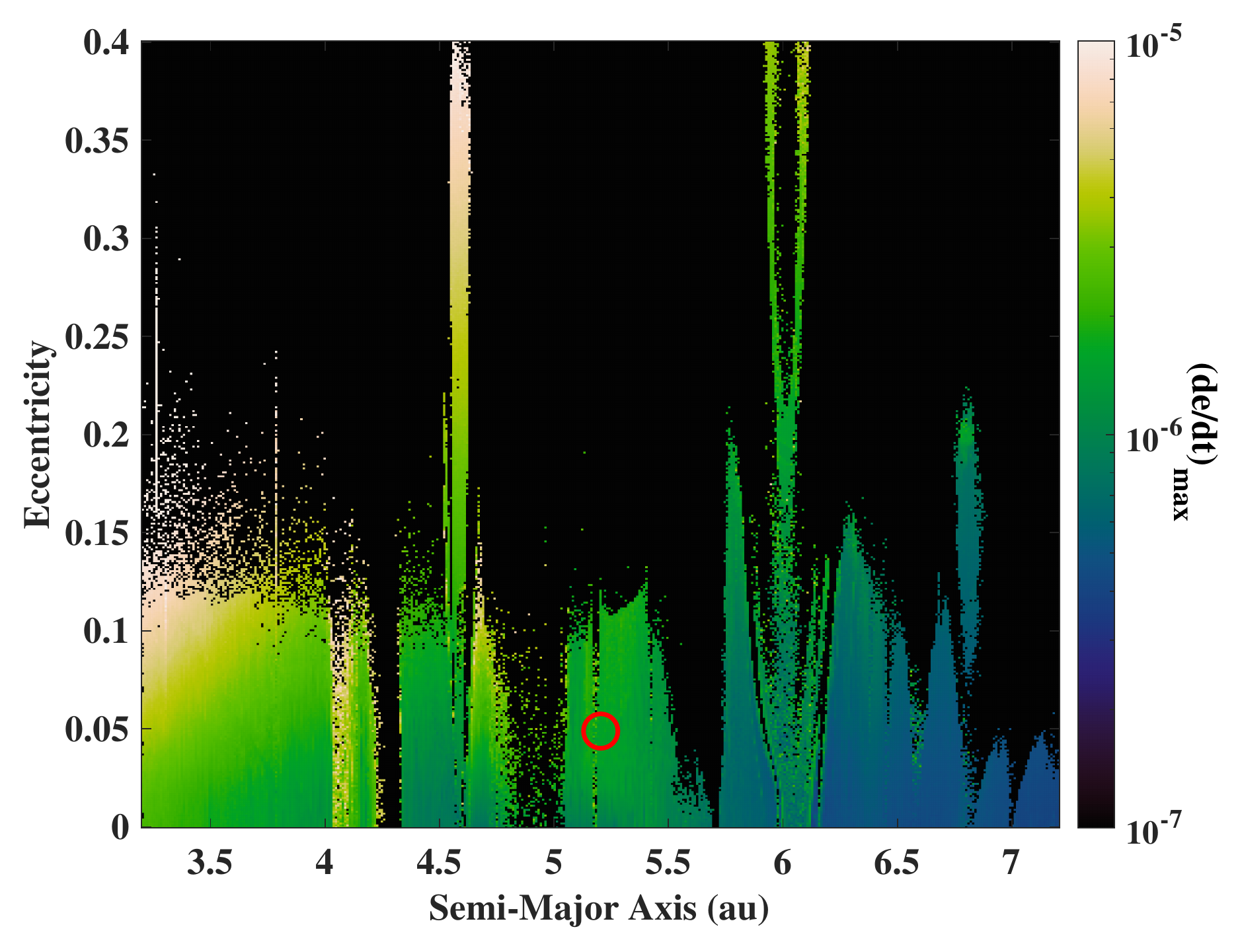}  \\
\end{tabular}
\caption{The variability of the Earth's orbital eccentricity, as a function of Jupiter's initial semi-major axis, $a$, and eccentricity $e$. The top left plot shows the maximum eccentricity obtained by Earth's orbit through the 10 Myr integrations. The top right plot shows the r.m.s. rate of change of Earth's eccentricity, plotted on a logarithmic scale. The lower two plots show the maximum rate of change of Earth's orbital eccentricity, plotted on a linear (left) and logarithmic (right) scale. For all plots, the black areas show those simulations for which the Solar system proved unstable, and so Earth's orbital variability was not assessed. The hollow red circle shows the location of Jupiter in the real Solar system.}
\label{ecc}
\end{figure*}

In Figure~\ref{ecc}, we show the impact of Jupiter's orbit on the time-evolution of Earth's orbital eccentricity. The regions in black in those plots are those for which the Solar system proved unstable in our simulations. The data plotted in the top-left panel of that figure reveals that, through most of the phase-space studied, Earth's maximum orbital eccentricity remains low, less than 0.1. However, there are small regions in which the Solar system is stable on the timescale of our integration, but Earth's eccentricity can be forced to larger values. In the main, these regions are those on the edge of stability, and it is quite possible that those systems would have become unstable with longer simulation time. Of perhaps more interest are the panels in Figure~\ref{ecc} showing the rate of change of Earth's orbital eccentricity. As Jupiter is moved closer to the Sun, both the mean and maximum rate of change of our planet's orbital eccentricity increase - a finding mirrored in the left-hand column of Figure~\ref{exemplar}.

\begin{figure}
\begin{tabular}{c}
  \includegraphics[width=0.5\textwidth]{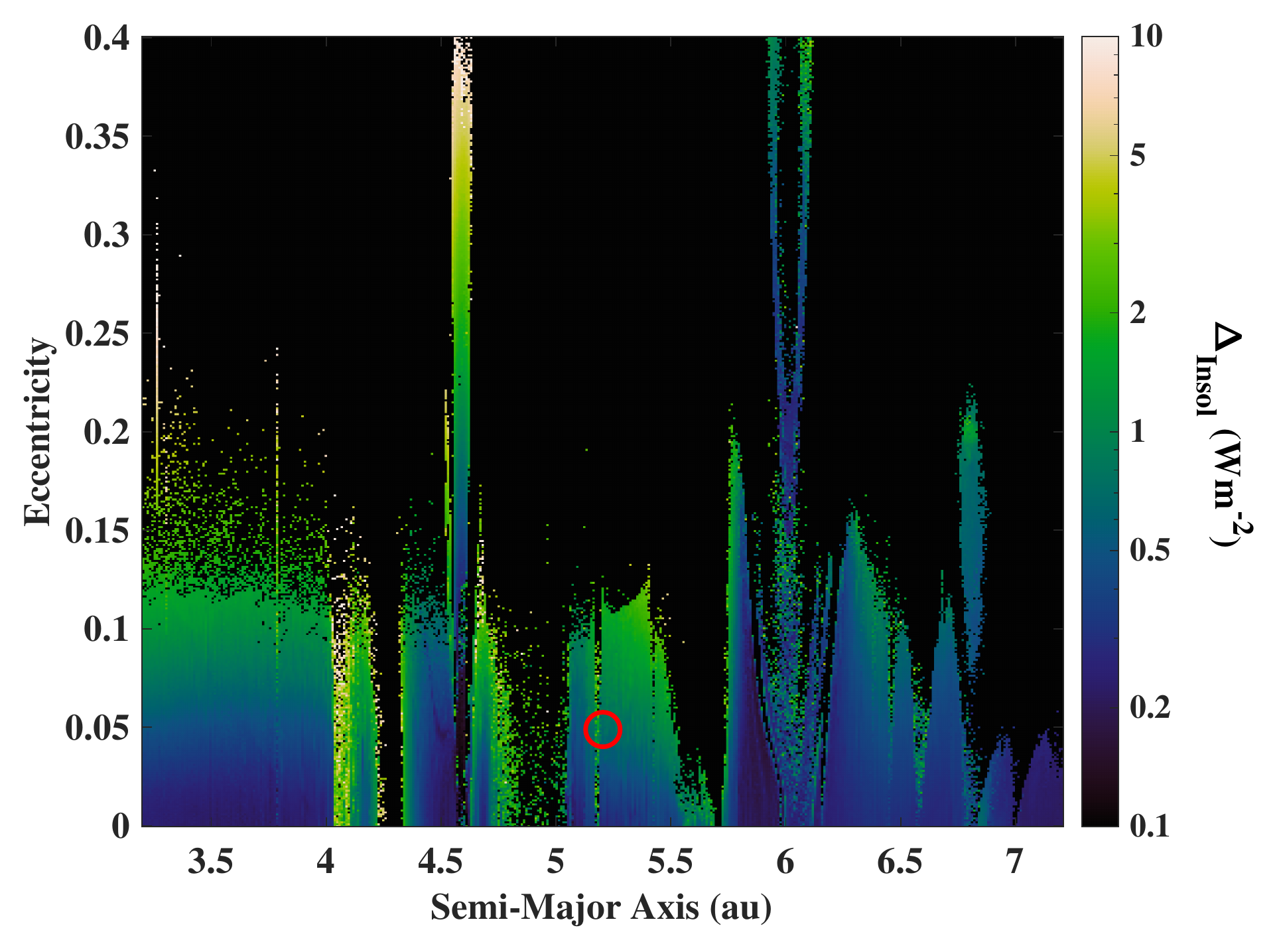}\\ 
\end{tabular}
\caption{The impact of Earth's orbital eccentricity on the annual mean insolation received by our planet, as a function of Jupiter's initial semi-major axis, $a$, and eccentricity, $e$. At each point in the plot, we show the difference between the maximum and minimum annual insolation Earth would experience in that scenario. The more eccentric Earth's orbit, the higher the annual insolation - and so this value represents the difference between the insolation when Earth's orbit is most eccentric and that when it is most circular.}
\label{insol}
\end{figure}

Given information on how Earth's orbital eccentricity changes with time, it is possible to quantify the degree to which the annual mean insolation on our planet changes. The higher Earth's orbital eccentricity, the higher the insolation averaged over the course of the year - as described in Equation~\ref{InsolationEquation} (where $\bar{F}_{annual}$ is the insolation averaged over the course of the year, $a$ is Earth's semi-major axis, $e$ the Earth's eccentricity, and $L_\odot$ is the luminosity of the Sun). For that reason, in Figure~\ref{insol}, we show the degree to which variations in Earth's orbital eccentricity would drive variations in annual mean insolation. We plot the difference between the highest and lowest values for annual mean insolation that would be experienced across the course of our integrations at a given location. Whilst the structure shown in these plots clearly follows the maximum eccentricity obtained by the Earth in our simulations, the non-linear response of insolation to changes in eccentricity can be clearly seen. As a point of reference, the current situation on Earth is that the difference in insolation between eccentricity maxima and minima is less than 0.5 Wm$^{-2}$. The most extreme scenarios tested in this work, for which Earth exhibits the greatest orbital eccentricities, would deliver changes in annual mean insolation that exceed those seen on Earth by more than an order of magnitude - a result that would have interesting implications for the planet's climate.

\begin{equation}
    \bar{F}_{annual} = {L_\odot \over{(16 \pi a^2 \sqrt{(1-e^2)})}}
\label{InsolationEquation}
\end{equation}

\begin{figure*}
\begin{tabular}{cc}
  \includegraphics[width=0.5\textwidth]{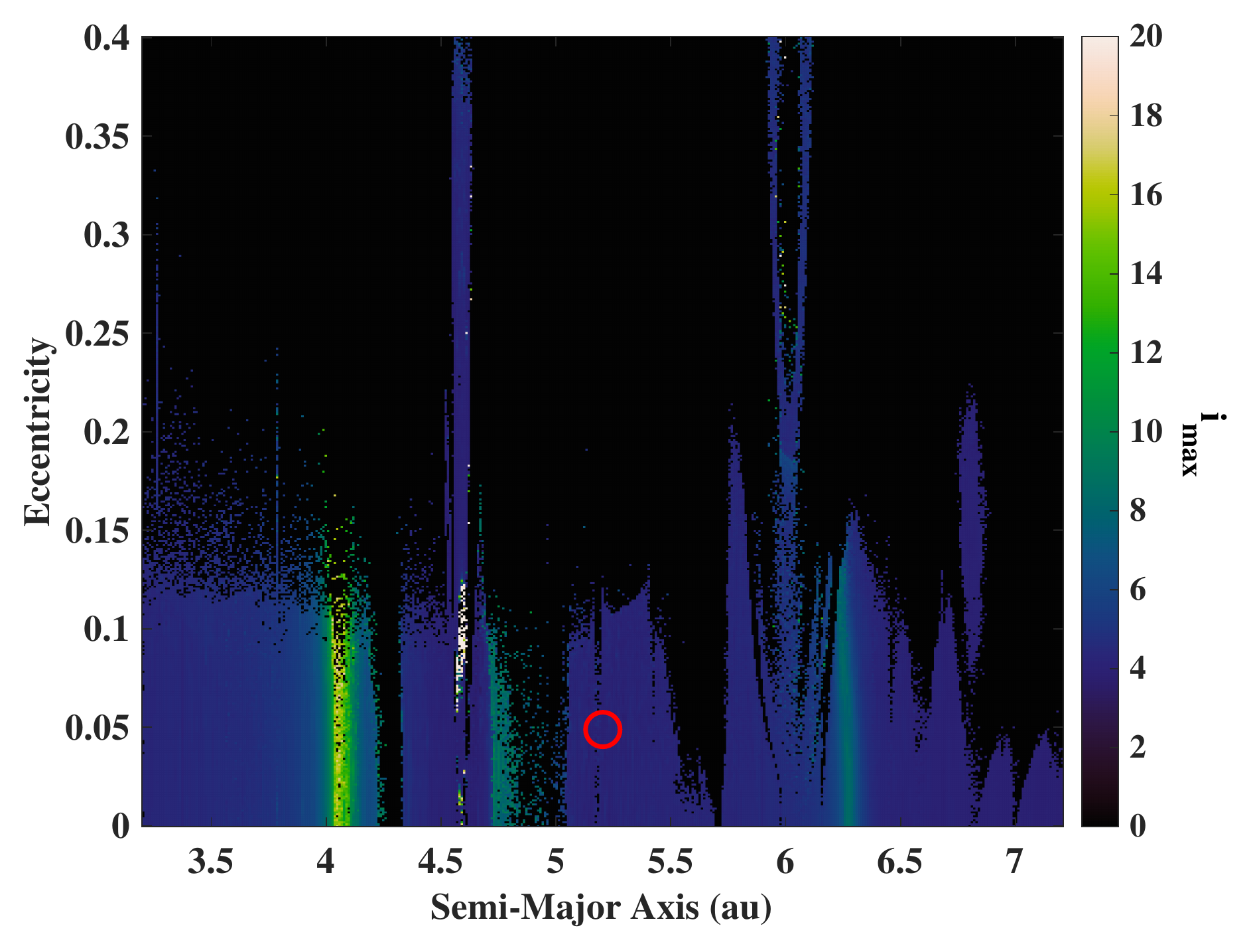} & \includegraphics[width=0.5\textwidth]{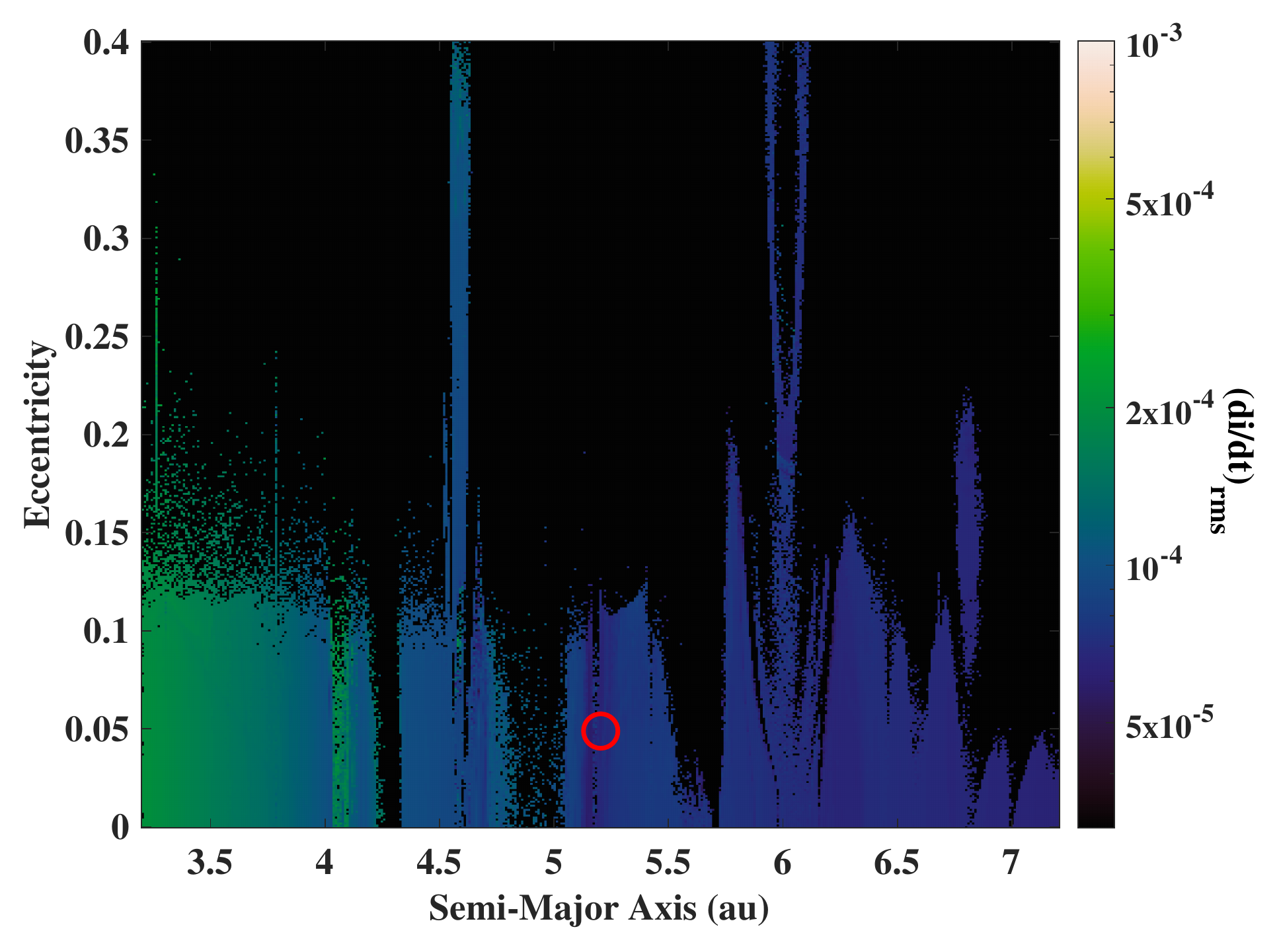}\\  \includegraphics[width=0.5 \textwidth]{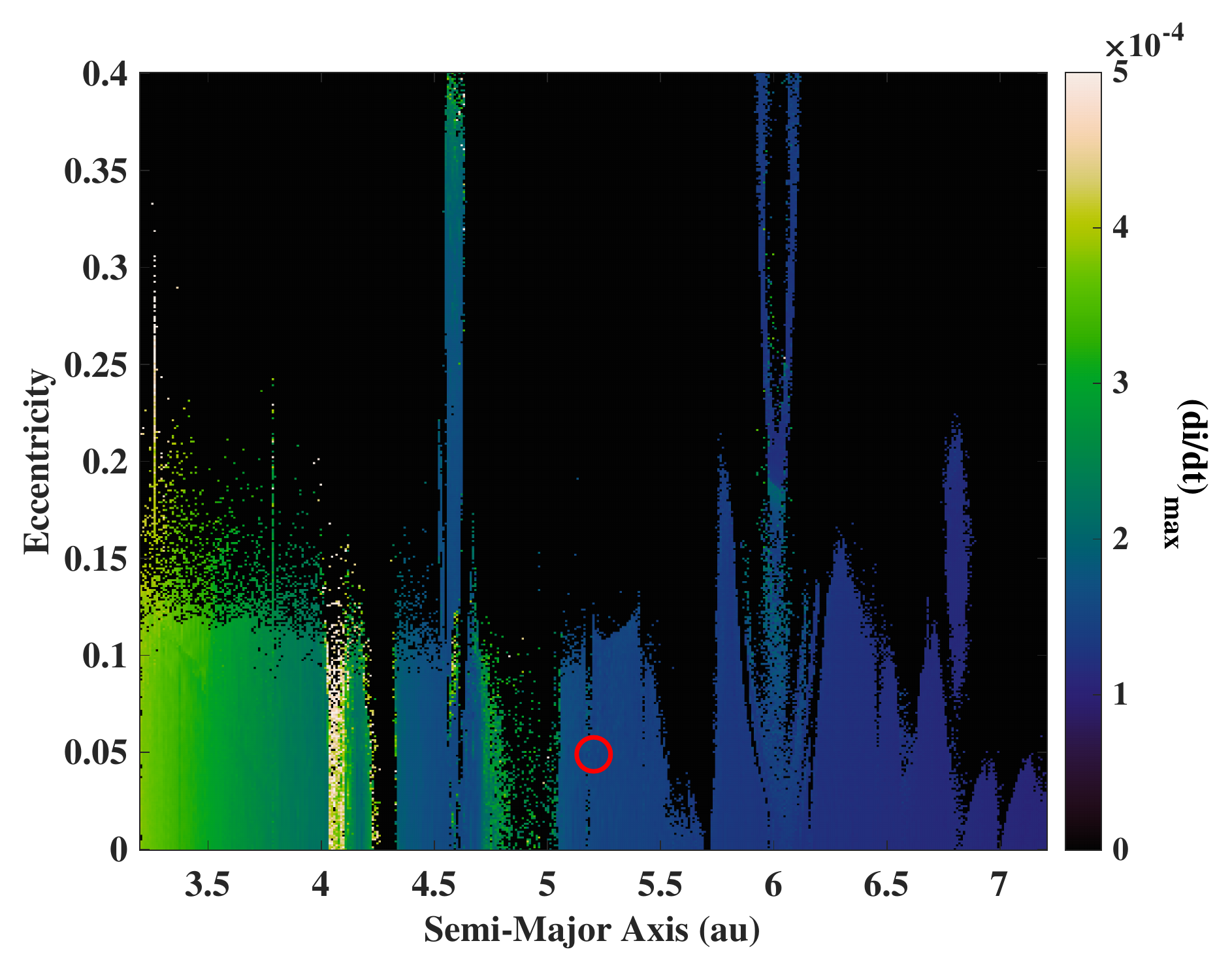} & \includegraphics[width=0.5 \textwidth]{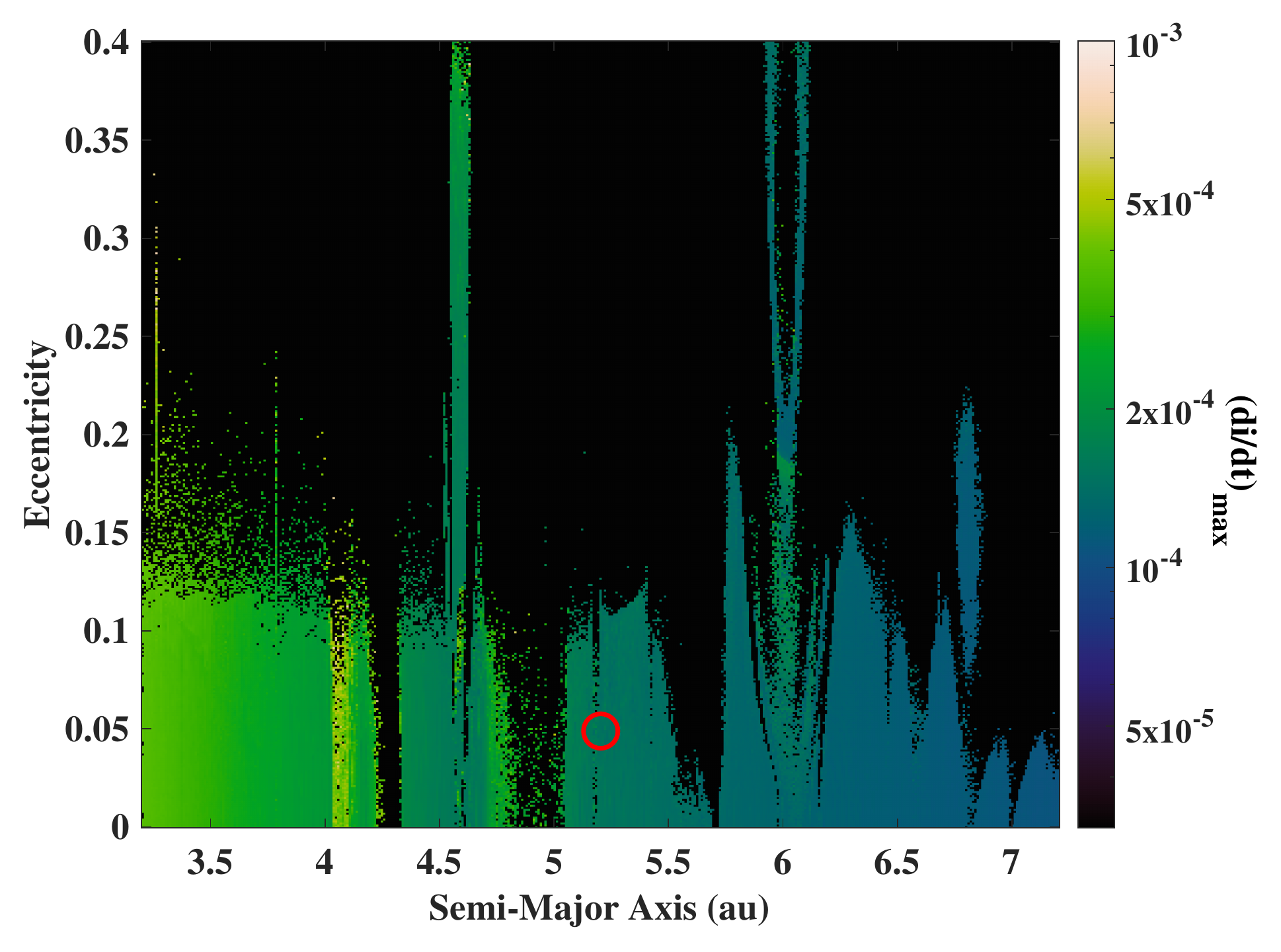}  \\
\end{tabular}
\caption{The variability of the Earth's orbital inclination, as a function of Jupiter's initial semi-major axis, $a$, and eccentricity $e$. The top left plot shows the maximum inclination obtained by Earth's orbit through the 10 Myr integrations. The top right plot shows the r.m.s. rate of change of Earth's inclination,  on a logarithmic scale. The lower two plots show the maximum rate of change of Earth's orbital inclination, plotted on a linear (left) and logarithmic (right) scale. For all plots, the black areas show those simulations for which the Solar system proved unstable, and so Earth's orbital variability was not assessed. The hollow red circle shows the location of Jupiter in the real Solar system.}
\label{inc}
\end{figure*}

The evolution of Earth's orbital inclination as a function of Jupiter's initial orbit is shown in Figure~\ref{inc}. In general, the maximum inclination to which Earth's orbit is excited remains low throughout the plot. The three exceptions to this are the regions just beyond 4~au, at around 4.75 and 6.3~au. At each of these locations, the Earth's orbit can experience significant inclination variability. The innermost of these regions also exhibits enhanced rates of inclination variability, as can be seen in the three other panels of Figure~\ref{inc}. The outermost band of increased $i_{max}$ values, however, shows no such feature - here, the rate of change of inclination is no greater than in the regions surrounding it.

These regions offer a cautionary tale for the assessment of potential exoplanet habitability -- just because the orbit of an Earth-like planet is stable, that does not mean that the oscillations in its orbit can not be relatively large. Such large excursions in the orbit of a dynamically stable planet could well have an impact on the evolution of its climate. To illustrate this, consider the impact of the planet's orbital inclination.

Given that the seasons are driven by the inclination of a planet's axis {\it with respect to the plane of its orbit}, rather than the absolute orientation of that axis in space, changing the inclination of a planet's orbit whilst the absolute orientation of the axial tilt remains fixed would result in an equivalent change in the tilt of the planet's axis {\it with respect to the plane of its orbit}. As a result, the extent of the polar circles would vary dramatically on a planet experiencing changes in orbital inclination greater than $10^\circ$, which would result in severe changes in seasonal insolation. Would such a planet still be habitable? Perhaps -- but it is probably fair to argue that such a world would be less promising as a target for the search for life than another that did not exhibit such extreme variability. 

The rates at which the Milankovitch-like oscillations occur may be key here. It has been suggested that rapid obliquity and eccentricity cycles could result in extreme seasonal variations that could trigger intense ice age cycles \citep[e.g.][]{dei18b}. On the other hand, it has been suggested that such rapid oscillations may suppress the ice-albedo feedback, and act to expand the outer edge of the HZ \citep[e.g.][]{armst14}. It seems plausible that each of those scenarios could occur. Which one is more likely to happen depends on the inertia of the system and the rate of change. Planets with a thicker atmosphere or larger ocean volume may, for example, respond more slowly to periodic alterations in the incoming stellar radiation and might therefore prove to be more resistant to oscillations in astronomical forcing than planets with thinner atmospheres and oceans \citep[e.g.][]{cow12}.

\begin{figure*}
\begin{tabular}{cc}
  \includegraphics[width=0.5\textwidth]{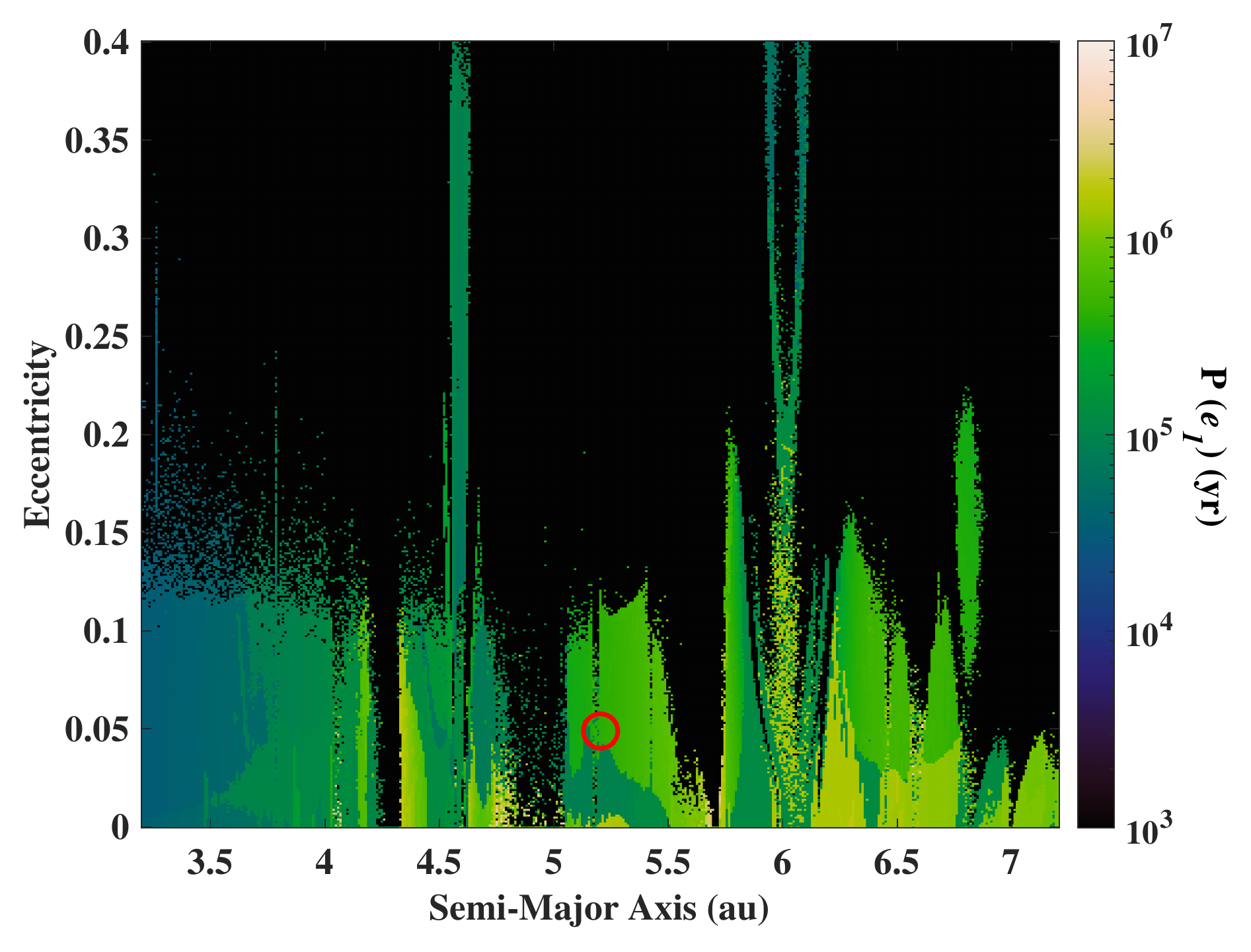} & \includegraphics[width=0.5\textwidth]{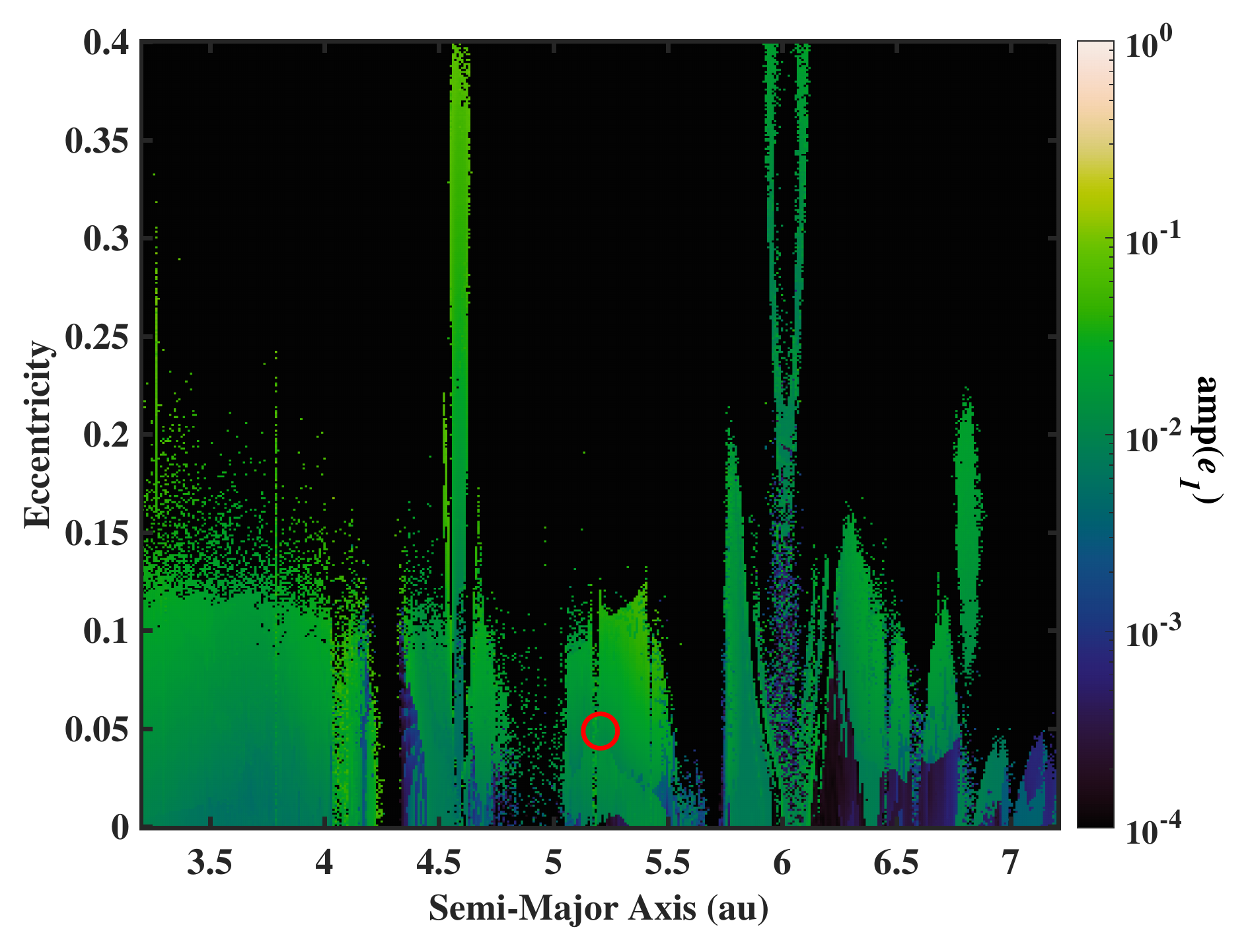}\\  \includegraphics[width=0.5 \textwidth]{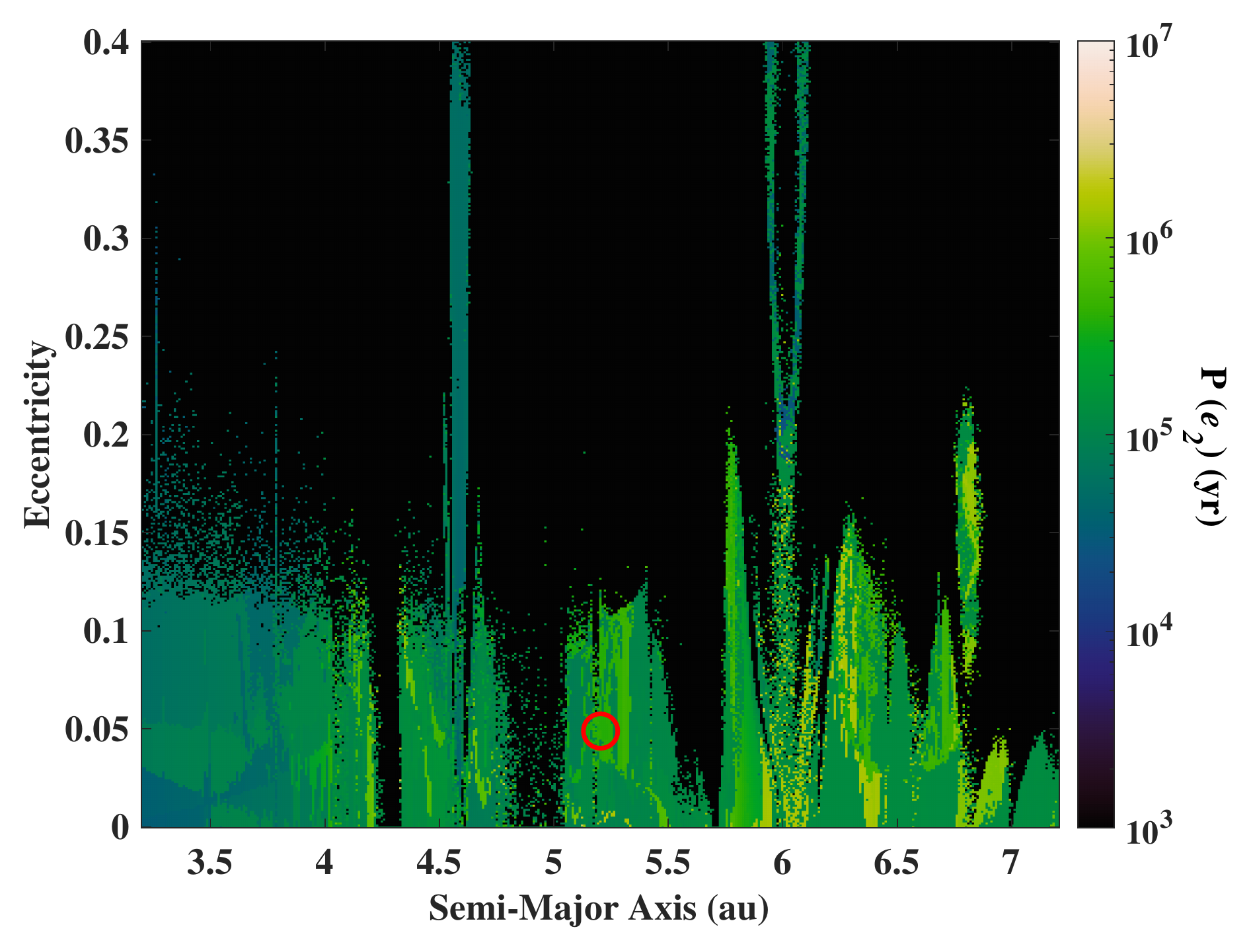} & \includegraphics[width=0.5 \textwidth]{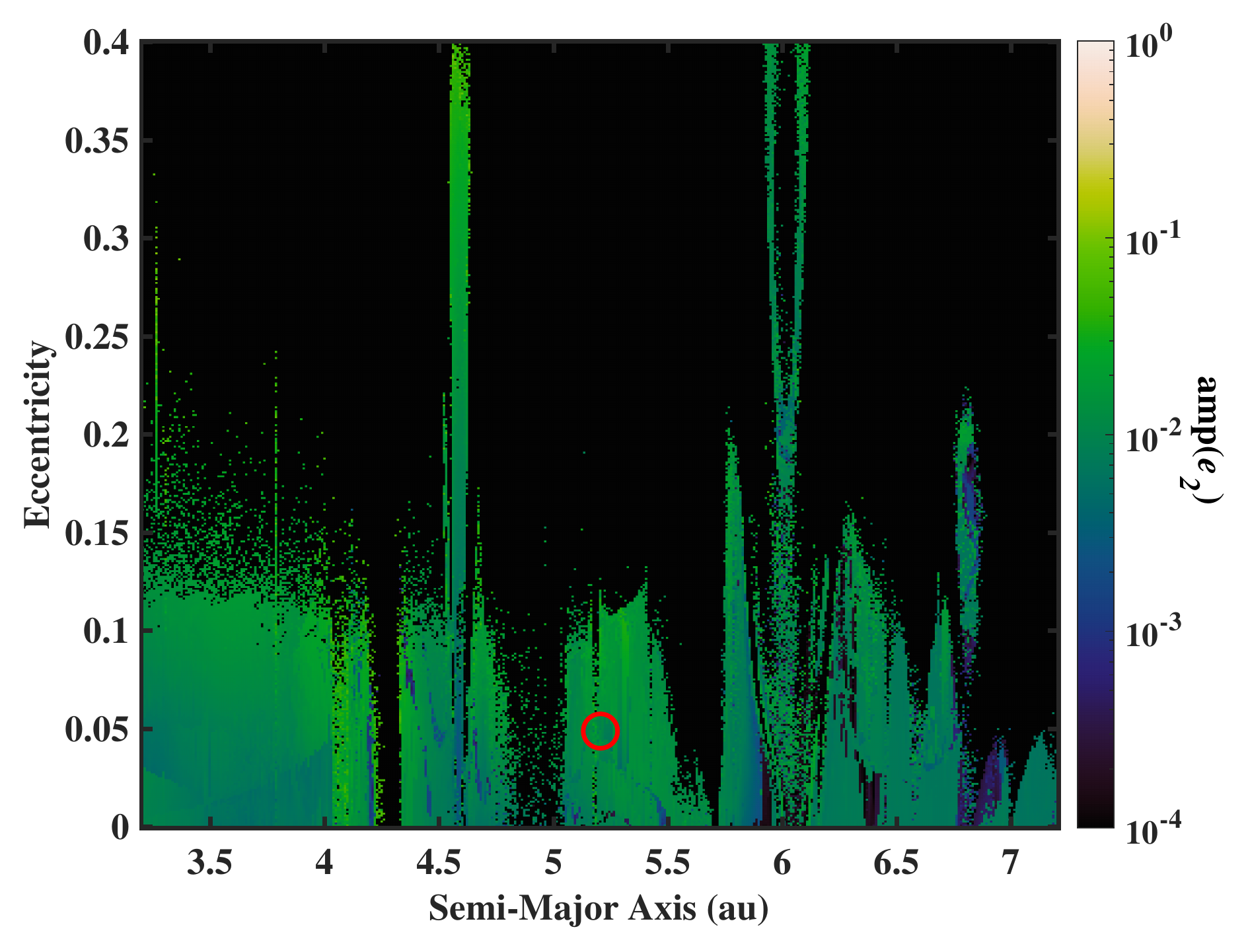}  \\
\end{tabular}
\caption{The period and amplitude of the two dominant oscillations in Earth's orbital eccentricity, as a function of Jupiter's initial semi-major axis, $a$, and eccentricity, $e$. The left-hand plots show the period (in years) of the two dominant frequencies, whilst the right-hand plots show their amplitudes.}
\label{ecc_2dw}
\end{figure*}

\begin{figure*}
\begin{tabular}{cc}
  \includegraphics[width=0.5\textwidth]{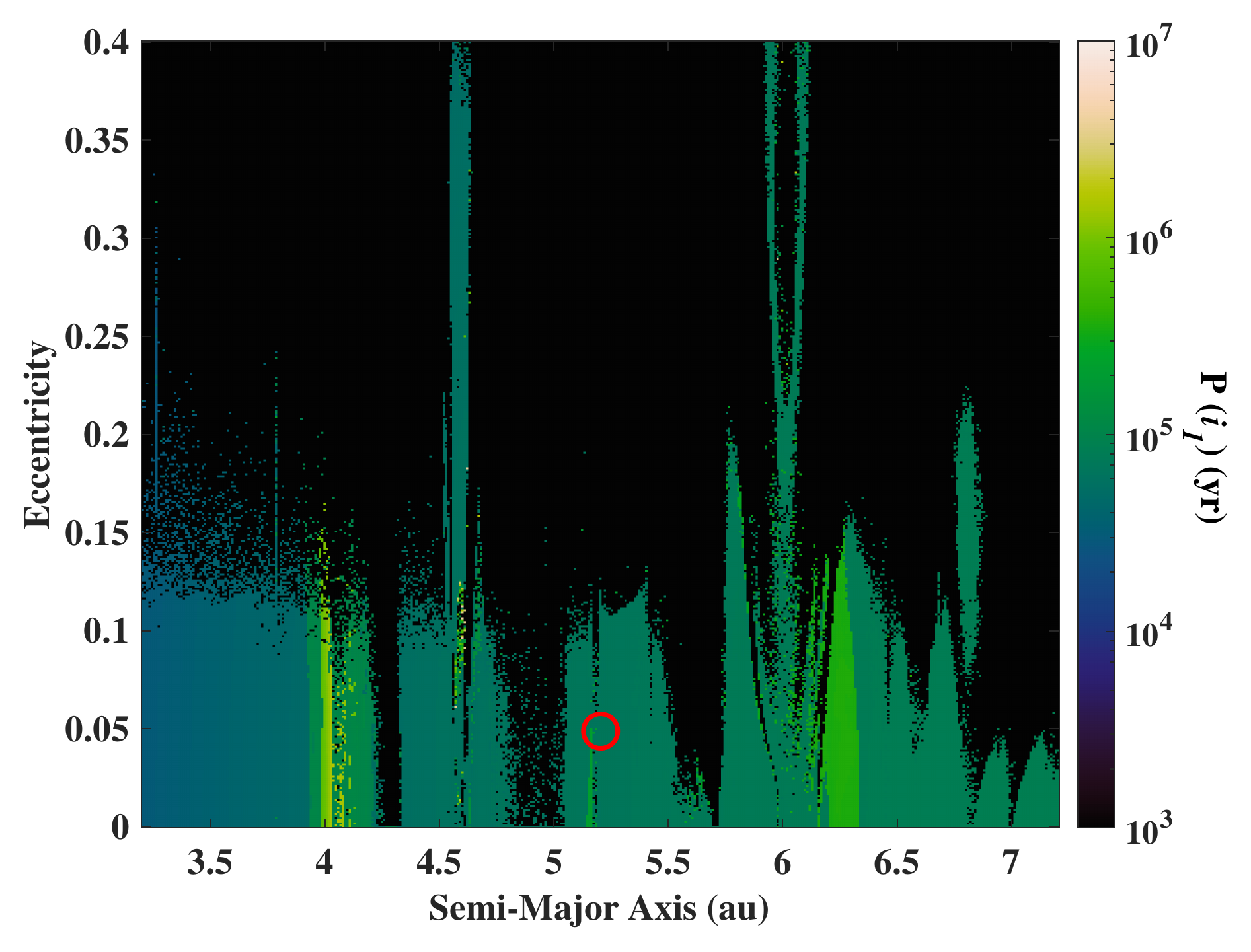} & \includegraphics[width=0.5\textwidth]{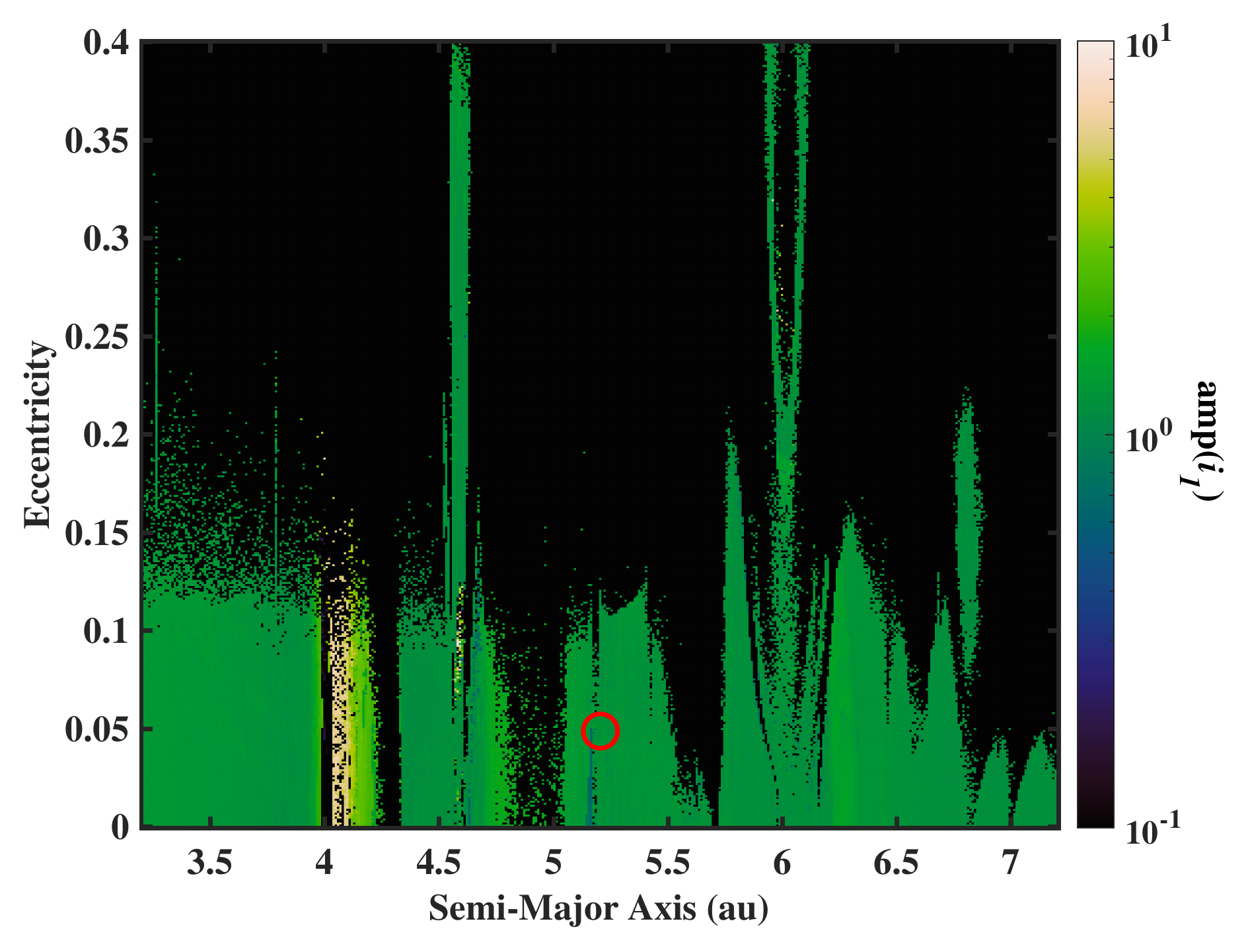}\\  \includegraphics[width=0.5 \textwidth]{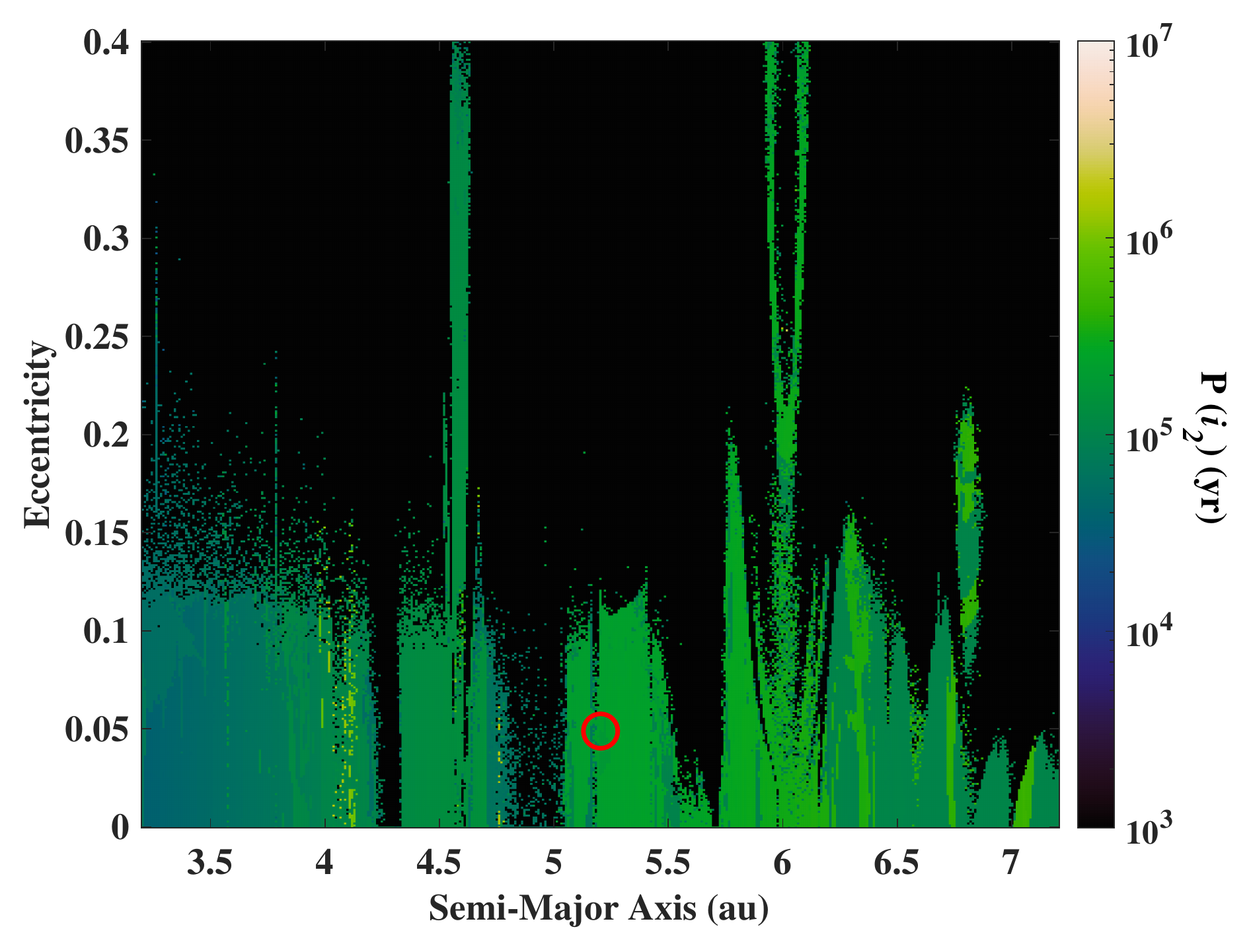} & \includegraphics[width=0.5 \textwidth]{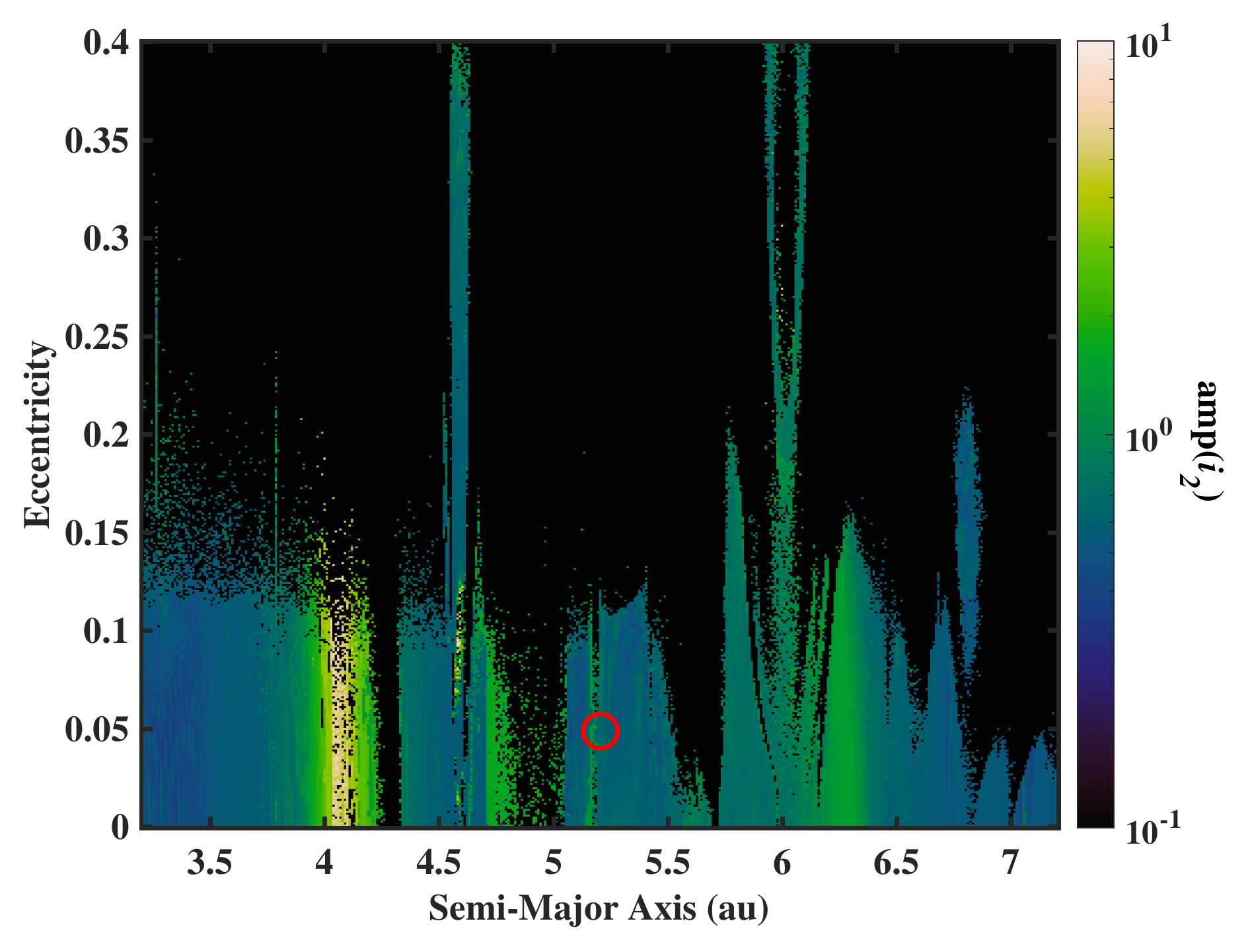}  \\
\end{tabular}
\caption{The period and amplitude of the two dominant oscillations in Earth's orbital inclination, as a function of Jupiter's initial semi-major axis, $a$, and eccentricity, $e$. The left-hand plots show the period (in years) of the two dominant frequencies, whilst the right-hand plots show their amplitudes, in degrees.}
\label{inc_2}
\end{figure*}

Figure~\ref{ecc_2dw} shows the period (left) and amplitude (right) of the two dominant oscillations in Earth's orbital eccentricity, with Figure~\ref{inc_2} showing the same information for the evolution of Earth's orbital inclination. In broad terms, Figures~\ref{ecc_2dw} and \ref{inc_2} show the same features that can be seen in Figure~\ref{fourierfigs}. Of particular interest is the 'phase-change' around 4~au, from a regime where both dominant periods occur on timescales $<$ 100 kyr to one that features both long- and short-period components.

\section{Discussion}
\label{discussion}

Variations in the incident flux received by a planet can affect its climate evolution and the evolution of its atmosphere, which both influence overall planetary habitability. For example, the notable effect of eccentricity on climate has been previously studied for a variety of scenarios and specific exoplanets \citep{wil02,kan12,way17,georg18}. The effects of obliquity and eccentricity on the incident flux for a planet have been quantified with application to exoplanets by \citet{kan17}. An exploration of how Milankovitch cycles affect exoplanet climates was recently undertaken by \citet{dei18a,dei18b}, showing the significant impact on obliquity variations and subsequent effect on planetary climate. 

This work is particularly important for the evaluation of the variation in flux for planets that lie within the HZ of their host stars \citep{kop13,kop14,kan16}. Given the diversity of exoplanet systems and their corresponding dynamics, it is possible and likely that combinations of orbital parameters exist which would produce stellar flux variations that render a HZ planet uninhabitable over its lifetime. It is also possible for the habitability of a planet to be time-dependent, and to be driven by periodic variations like Milankovitch cycles. Studies such as ours therefore play an important role -- they allow the current dynamical state of the system to be assessed, but can also be used to examine the likelihood that the planet has remained 'habitable' on astronomically-long timescales. 
%Such extended epochs of habitability are necessary for the development of self-replicating molecules and the origination of life. HZ exoplanets exhibiting histories with higher probabilities of habitable epochs ought to be ranked higher in importance for target selection. 

In addition to studying the short-term habitability of a system, simulations such as ours can also reveal the degree to which a given Milankovitch regime is robust against the migration of the planets in that system. For example, our simulations suggest that, were Jupiter to migrate just a short distance to $\sim$ 5~au from the Sun, the Solar system could become catastrophically unstable. Such migration-driven instability is not a new concept in the narrative of Solar system evolution, having been invoked in the past to explain the Late Heavy Bombardment of the inner Solar system \citep[e.g.][]{LHB1,LHB2,LHB3}.

Even on smaller scales, the fine structure visible in the plots of Earth's orbital evolution suggests that even relatively small-scale migration could cause marked shifts in the potential habitability of a planet. Given that Jupiter is continually ejecting cometary and asteroidal material from the Solar system \citep[e.g.][]{Horner03,FoF1,FoF2,Grazier2,Grazier3}, it must still be undergoing a very gradual inward migration. The same will no doubt be true of planets orbiting other stars -- nothing is truly static on astronomical timescales. 

By performing simulations like those presented herein for potentially habitable exoEarths, it may even be possible to identify those that might have experienced catastrophic or chaotic Milankovitch cycles in the past. Whether such periods of instability would be deleterious or beneficial to the development of detectable life is still open to debate. Compare, for example, the impact of the ancient episodes of 'Snowball Earth', which have been suggested as potentially contributing to the explosion of new species during the Neoproterozoic \citep[e.g.][]{Hoffman,kirschvink}, to the effect that such an event would have on life at the modern era \citep[which would likely be catastrophic, given that even the relatively minor recent glaciations have had a marked effect on the Earth's biodiversity; e.g.][]{wil93,hew00,hew04,snid13,Avian15}. Despite the complexity of such situations, the ability to identify such 'edge cases' may help focus the choice of the most promising targets in the search for life, and so should definitely be considered in future work.

\section{Implications for the Rare Earth Hypothesis}

In this work, we have solely considered the impact of the orbital parameters (eccentricity, inclination, apsidal precession) that contribute to variations in the mean annual flux received at the top of a planet's atmosphere. Elements in the axial group (obliquity and axial precession) clearly play an important role in determining the spatial distribution of incident flux on a planet, but these variables cannot be measured for exoplanets at the current time. Observations to determine those angles will be as challenging as those required to search for evidence of life on the planets considered. As our work is intended to help guide the selection of the targets for such observations, our focus lies on those parameters that we might reasonably measure in the near future.

Nonetheless, it is interesting to briefly view our results in the context of the real Solar system. In the past, it has been argued that the Earth is unusually favourable for the development of life -- a core tenet of the `Rare Earth' hypothesis \citep[e.g.][]{rareearth,waltearth}. In that light, it is interesting to consider the degree to which our Earth is unusual in the context of our simulations. Are the orbital variations experienced by our planet unusual or typical, when compared to our ensemble of `alternate Earths'? We stress that such a study is purely illustrative, since our results take no account of the impact on our planet's climate from elements in the axial group. 

\begin{figure*}
\begin{tabular}{c}
  \includegraphics[width=1.0\textwidth]{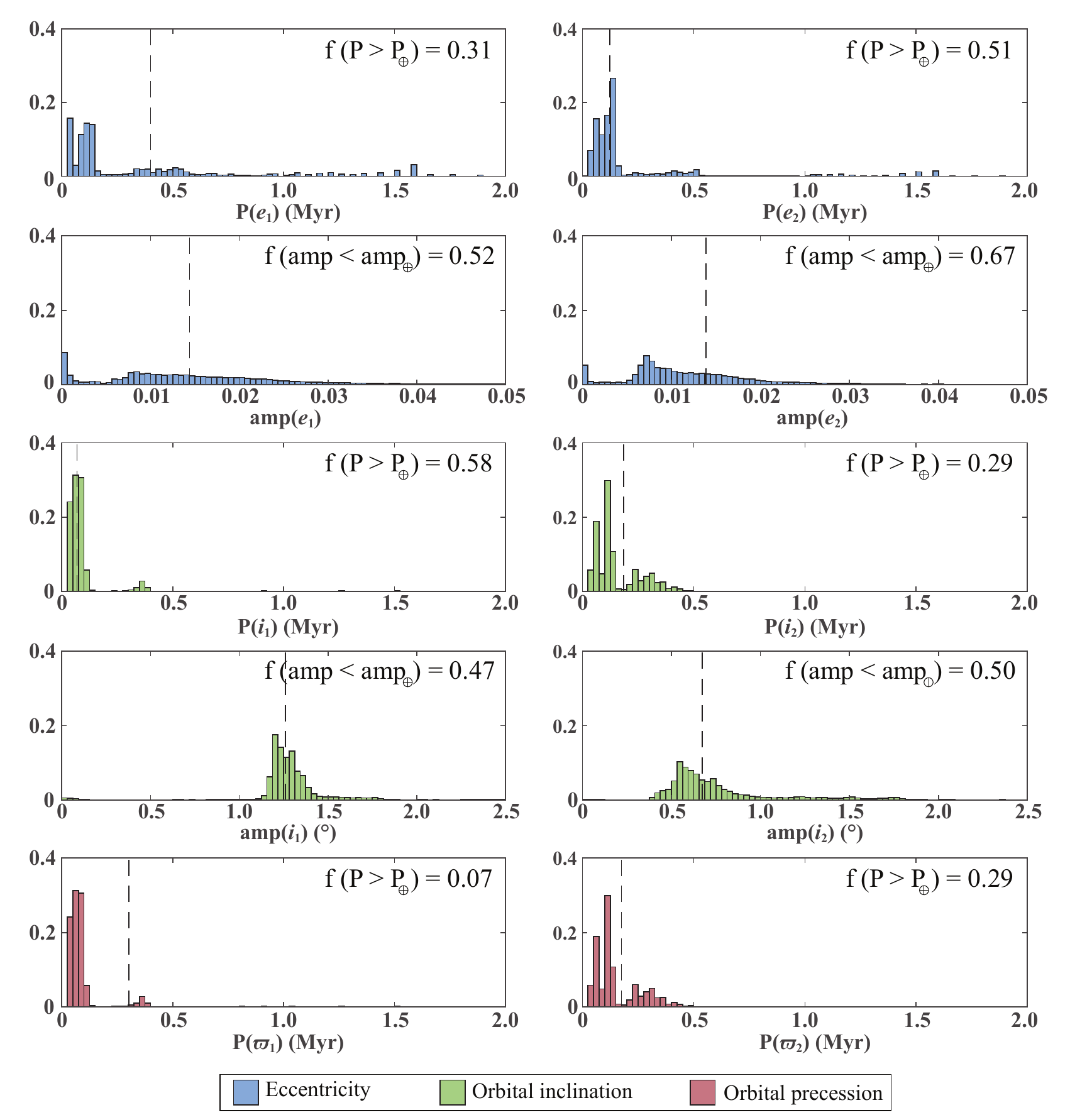} \\ 
\end{tabular}
\caption{Histograms showing the fraction of simulations in which the Earth's orbital evolution exhibited oscillations of a given amplitude and period. The left-hand column shows the dominant oscillation for each variable, the right-hand column showing the second strongest oscillations. The first two rows (with data plotted in blue) show the cyclic variations in Earth's orbital eccentricity, with the period (row 1) and associated amplitude (row 2). The third and fourth rows (in green) show the same information for the Earth's orbital inclination, and the fifth row (in red) show variations in the longitude of Earth's perihelion. In each plot, the vertical dashed lines show the location at which the `true Earth', in our Solar system, would fall on the plots. The numerical values detailed in the boxes themselves denote the fraction of the total ensemble of stable simulations for which the period of the oscillations is longer than that for the `true Earth', or the fraction with oscillations whose amplitude is smaller than that seen in our Solar system. When compared with our ensemble of stable simulations, the real Earth's periodic orbital oscillations are fairly typical - neither unusually large nor unusually small.}
\label{histo}
\end{figure*}

To consider what fraction of the stable simulations would be equally or more clement than our Solar system (with low amplitude and low frequency oscillations), we calculated the distribution of frequencies and amplitudes for each of the orbital parameters that impact the Milankovitch cycles. The results of this analysis are shown in Figure~\ref{histo}. The two uppermost panels show the distribution of the periods and amplitudes of the two strongest periodic oscillations in Earth's orbital eccentricity, across our runs. It becomes apparent that the periods of oscillation span a broad range. In total, the fraction of Earths whose dominant eccentricity cycle had a period longer than Earth's 400 kyr cycle was 31\%, whilst 51\% of simulations featured a secondary eccentricity cycle with period greater than Earth's 100 kyr periodicity. The amplitude of the dominant eccentricity cycle was smaller than that seen in our Solar system in 52\% of cases, whilst the secondary cycle amplitudes were smaller than those we experience in 67\% of cases. In other words, when it comes to the variability of our orbital eccentricity, the Earth seems unremarkable.

In a similar fashion, we can examine the evolution of Earth's orbital inclination and the longitude of our planet's perihelion in the context of the ensemble of stable runs. Once again, the Earth's orbital evolution is relatively typical of the ensemble. 58\% of systems featured primary inclination variability on timescales longer than that experienced by the Earth, with 29\% showing secondary inclination periodicity on longer timescales. 47\% of systems feature primary inclination oscillations smaller than those for our planet, with 50\% of systems having smaller secondary oscillations. 

The rate at which Earth's longitude of perihelion precesses is at the more sedate end of those seen in our simulations. In total, 80\% of stable scenarios featured `Earths' whose dominant perihelion precession rate was more rapid than seen for our planet, with 67\% exhibiting more rapid precession for the second strongest periodicity.

Finally, we note for posterity the extremes in the variability observed across our dynamical stable ensemble of simulations. Across those 41,652 runs, the largest eccentricity obtained by the Earth was 0.415 in the scenario where Jupiter was initially located at a semi-major axis of 4.04~au and has an eccentricity of 0.168. Such an extreme eccentricity value for the Earth would result in an increase in the annual mean solar flux of ca. 10\%, or 24~Wm$^{-2}$ when also accounting for our planet's modern albedo -- a significant change compared to the 0.2\% variation received by Earth over the eccentricity time scales associated with the glacial-interglacial fluctuations.

To put this into the context of the Earth's modern climate sensitivity, we note that the estimated radiative forcing that would be caused by a doubling of atmospheric CO$_2$ is 3-4~Wm$^{-2}$, a value which takes into account various climate processes such as atmospheric water vapour, cloud, and lapse rate feedbacks \citep[]{andrews2012,huang}. An increase of 24~Wm$^{-2}$ would be comparable to the rise in global temperature that would result from eight doublings of atmospheric CO$_2$. 

It is conventionally assumed that global temperatures increase by 0.8~K per Wm$^{-2}$. An increase of 24~Wm$^{-2}$ would therefore equate to a global warming of approximately 20~K. Similar estimates have been obtained in previous climate modelling studies that assessed temperature variability for Earth-like planets on orbits of various eccentricities \citep[]{wil02,dres2010}. Such major global temperature changes are unprecedented in the Phanerozoic era of Earth (past 500 Myr), based on estimates from paleoclimate records \citep[e.g.][]{royer2004,hansen2013}.

%%%%%%%%%%%%%%%%%%%%%%%%%%%%%%%%%%%%%%%%%%%%%%%%%%%%%%%%%%%%%%%%%%%%

\section{Conclusions}
\label{conclusions}

In the coming years, exoplanet experiments will produce an improved sensitivity to Earth-sized planets orbiting within the HZ of solar-type stars. Such planets will become the targets in our efforts to search for evidence of life beyond the Solar system. However, such observations will be immensely challenging and, in the early stages of that search, we will only be able to study a small subset of the discovered planets. It is therefore vital that we prioritize which of those planets are the most promising targets for intensive follow-up observations. Such prioritization will consider many factors that come together to render one planet more, or less, habitable than another \citep[e.g.][]{HabRev}. An important example of such a factor is the orbital dynamics of the planet and the subsequent effects on the planetary climate.

In this work, we detail the results of a large suite of $n$-body simulations designed to examine the influence of Jupiter on the orbital cycles experienced by the Earth. We systematically varied the initial orbit of Jupiter across a region spanning $\pm$ 2~au in semi-major axis around the orbit of Jupiter in our Solar system. At each unique semi-major axis we tested, we varied the initial orbital eccentricity of the giant planet in the range 0.0-0.4. This yielded a grid of almost 160,000 unique variants of our Solar system - each of which featured the other seven planets (Mercury, Venus, Earth, Mars, Saturn, Uranus and Neptune) moving on identical orbits - those that they occupy in the Solar system. We simulated the evolution of these modified Solar systems over a period of 10 Myr, and examined how Earth's orbit subsequently varied over time.

Our results reveal the sensitivity of the Solar system's stability to Jupiter's orbit - with some $\sim74\%$ of the variant systems proving catastrophically unstable within the ten million years of our integrations. For the subset that proved stable, we found that both the periods and amplitudes of the oscillations in Earth's orbital elements varied markedly as a function of Jupiter's initial orbital elements. When Jupiter began on an orbit closer to the Sun, the periodicity of Earth's orbital element variation was typically shorter than when Jupiter was more distant. Simultaneously, the amplitude of the Earth's orbital cycles varied as the giant planet was moved through the Solar system -- with some stable Solar system variants featuring oscillations in Earth's orbital inclination that approached, or even exceeded, ten degrees. 

Our work highlights the degree to which small changes in the architecture of a planetary system can drive large variations in the Milankovitch cycles that would be experienced by any potentially habitable planets therein. It will therefore be critically important to accurately determine the orbital elements for all planets within the system, in order to assess the scale and frequency of their orbital cycles so that the impact of Milankovitch forcing on terrestrial planets can be considered.

When the amplitudes and periods for the dominant cycles in Earth's orbital eccentricity, inclination, and the precession of our planet's perihelion are compared with the results across our ensemble of stable solutions, we find that our planet's orbital behaviour is remarkably unremarkable. The variations in the Earth's orbital parameters are neither unusually fast nor unusually slow, nor are they unusually large or small. As such, at least when it comes to the orbital Milankovitch cycles, it seems that the central tenet of the `Rare Earth' hypothesis does not hold true. The Earth is not unusual - and so it may be that planets with similar Milankovitch cycles are common in the cosmos.

The transit- and radial velocity follow-up of {\it TESS} planets will likely reveal a number of other planets in the same systems, which will enable studies such as the one described here to be carried out for those new systems. Beyond simply allowing us to compare the potential habitability of those planets, such studies will form a fascinating complement to the atmospheric observations carried out by {\it JWST} and other facilities. The potential combination of our being able to study both the orbital architectures and atmospheric properties for these {\it TESS} planets will offer an unprecedented opportunity to directly examine the influence of orbital dynamics on planetary atmospheres outside of our Solar system.

%%%%%%%%%%%%%%%%%%%%%%%%%%%%%%%%%%%%%%%%%%%%%%%%%%%%%%%%%%%%%%%%%%%%

\section*{Acknowledgements}

This research has made use of the Habitable Zone Gallery at
hzgallery.org. This research has also made use of the NASA Exoplanet
Archive, which is operated by the California Institute of Technology,
under contract with the National Aeronautics and Space Administration
under the Exoplanet Exploration Program. The results reported herein
benefited from collaborations and/or information exchange within
NASA's Nexus for Exoplanet System Science (NExSS) research
coordination network sponsored by NASA's Science Mission Directorate. PV and SKT were supported by a Heising-Simons Foundation award while this research was ongoing. The authors wish to express their gratitude to the anonymous referee, whose suggestions helped to greatly improve the flow and clarity of our work.

\software{\textsc{Mercury}\citep{Mercury}}
\software{\textsc{Astrochron R package}}

%%%%%%%%%%%%%%%%%%%%%%%%%%%%%%%%%%%%%%%%%%%%%%%%%%%%%%%%%%%%%%%%%%%%


\begin{thebibliography}{}

\bibitem[Adams(2010)]{adams2010} Adams, F.~C.\ 2010, Annual Review of Astronomy and Astrophysics, 48, 47
\bibitem[Agnew et al.(2017)]{Agnew17} Agnew, M.~T., Maddison, S.~T., Thilliez, E., et al.\ 2017, \mnras, 471, 4494
\bibitem[Agnew et al.(2018a)]{Agnew18a} Agnew, M.~T., Maddison, S.~T., \& Horner, J.\ 2018, \mnras, 477, 3646
\bibitem[Agnew et al.(2018b)]{Agnew18b} Agnew, M.~T., Maddison, S.~T., \& Horner, J.\ 2018, \mnras, 481, 4680
\bibitem[Agnew et al.(2019)]{Agnew19} Agnew, M.~T., Maddison, S.~T., Horner, J., et al.\ 2019, \mnras, 485, 4703
\bibitem[Albrecht et al.(2012)]{albrecht12} Albrecht, S., Winn, J.~N., Johnson, J.~A., et al.\ 2012, \apj, 757, 18
\bibitem[Andrews et al.(2012)]{andrews2012}Andrews, T., Gregory, J.M., Webb, M.J. \& Taylor, K.E., 2012. Geophysical Research Letters, 39(9).
\bibitem[Armstrong et al.(2014)]{armst14} Armstrong, J.C., Barnes, R., Domagal-Goldman, S., et al. 2014, Astrobiology, 14, 277
\bibitem[Arya et al.(2017)]{ary17} Arya, M., Webb, D., McGown, J., et al. 2017, SPIE, 10400, 1C
\bibitem[Bailes et al.(2011)]{psr1719} Bailes, M., Bates, S.~D., Bhalerao, V., et al.\ 2011, Science, 333, 1717 
\bibitem[Barnes et al.(1983)]{barnes83}Barnes, R.T.H., Hide, R., White, A.A. \& Wilson, C.A., 1983, Proceedings of the Royal Society of London. A. Mathematical and Physical Sciences, 387, 1792, pp.31-73.
\bibitem[Barnes et al.(2015)]{Barnes15} Barnes, R., Meadows, V.~S., \& Evans, N.\ 2015, \apj, 814, 91 
\bibitem[Barstow \& Irwin(2016)]{barhab} Barstow, J.~K., \& Irwin, P.~G.~J.\ 2016, \mnras, 461, L92 
\bibitem[Batalha et al.(2013)]{Batalha13} Batalha, N.~M., Rowe, J.~F., Bryson, S.~T., et al.\ 2013, \apjs, 204, 24 
\bibitem[Batygin(2015)]{batygin2015} Batygin, K.\ 2015, \mnras, 451, 2589
\bibitem[Bennett(1990)]{bennett} Bennett, K.D. 1990, Paleobiology, 16, 11
\bibitem[Beichman et al.(2014)]{beijwst} Beichman, C., Benneke, B., Knutson, H., et al.\ 2014, \pasp, 126, 1134 
\bibitem[Berger(1978)]{berger78} Berger, A.\ 1978, AJ, J. Atmos. Sci., 35, 2362
\bibitem[Berger and Loutre(1994)]{berger94} Berger, A. \& Loutre, M.F.\ 1994, Long-Term Climatic Variations, pp. 107-151. Springer, Berlin, Heidelberg 
\bibitem[Berger et al.(1999)]{berger1999}Berger, A., Li, X.S. \& Loutre, M.F., 1999, Quaternary Science Reviews, 18, 1, pp.1-11.
\bibitem[Borucki et al.(2010)]{Borucki10} Borucki, W.~J., Koch, D., Basri, G., et al.\ 2010, Science, 327, 977 
\bibitem[Bouchy et al.(2005)]{Bouchy05} Bouchy, F., Udry, S., Mayor, M., et al.\ 2005, \aap, 444, L15 
\bibitem[Burgasser et al.(2010)]{ross458c} Burgasser, A.~J., Simcoe, R.~A., Bochanski, J.~J., et al.\ 2010, \apj, 725, 1405 
\bibitem[Campbell et al.(1988)]{GammaCeph} Campbell, B., Walker, G.~A.~H., \& Yang, S.\ 1988, \apj, 331, 902 
\bibitem[Chambers(1999)]{Mercury} Chambers, J.~E.\ 1999, \mnras, 304, 793 
\bibitem[Charbonneau et al.(2009)]{SE1} Charbonneau, D., Berta, Z.~K., Irwin, J., et al.\ 2009, \nat, 462, 891 
\bibitem[Clemens \& Tiedemann(1997)]{clem}Clemens, S.C. \& Tiedemann, R. 1997, \nat, 385, 6619
\bibitem[Correia, \& Laskar(2009)]{Correia} Correia, A.~C.~M., \& Laskar, J.\ 2009, \icarus, 201, 1.
\bibitem[Corsetti et al.(2006)]{Corsetti} Corsetti, F.A., Olcott, A.N., Bakermans, C.,\ 2006, Paleogeography Paleoclimatology Paleoecology, 232, 114.
\bibitem[Cowan et al.(2012)]{cow12}Cowan, N.B., Voigt, A. \& Abbot, D.S., 2012, The Astrophysical Journal, 757(1), p.80.
\bibitem[Cronin(2009)]{Cronin}Cronin, T. M., 2009, Paleoclimates: understanding climate change past and present, Columbia University Press
\bibitem[Cuntz \& Guinan(2016)]{Cuntz16} Cuntz, M., \& Guinan, E.~F.\ 2016, \apj, 827, 79 
\bibitem[Curry et al.(1995)]{curry} Curry, J.A., Schramm, J.L. and Ebert, E.E. \ 1995, Journal of Climate, 8, 240
\bibitem[Dehant et al.(1990)]{dehant90}Dehant, V., Loutre, M.F. \& Berger, A., 1990, Journal of Geophysical Research: Atmospheres, 95, D6, pp.7573-7578.
\bibitem[Deitrick et al.(2018a)]{dei18a} Deitrick, R., Barnes, R., Quinn, T.R., et al. 2018a, AJ, 155, 60
\bibitem[Deitrick et al.(2018b)]{dei18b} Deitrick, R., Barnes, R., Quinn, T.R., et al. 2018b, AJ, 155, 266
\bibitem[Des Marais et al.(2002)]{Des02} Des Marais, D.~J., Harwit, M.~O., Jucks, K.~W., et al.\ 2002, Astrobiology, 2, 153 
\bibitem[Dressing et al.(2010)]{dres2010} Dressing, C.~D., Spiegel, D.~S., Scharf, C.~A., Menou, K., \& Raymond, S.~N.\ 2010, \apj, 721, 1295 
\bibitem[Dressing \& Charbonneau(2013)]{dress13} Dressing, C.~D., \& Charbonneau, D.\ 2013, \apj, 767, 95
\bibitem[Duncan, \& Quinn(1993)]{duncan1993} Duncan, M.~J., \& Quinn, T.\ 1993, Annual Review of Astronomy and Astrophysics, 31, 265
\bibitem[Eggl et al.(2013)]{egg13} Eggl, S., Haghighipour, N., Pilat-Lohinger, E. 2013, ApJ, 764, 130
\bibitem[Encrenaz(2008)]{encrenaz2008} Encrenaz, T.\ 2008, Annual Review of Astronomy and Astrophysics, 46, 57
\bibitem[Fang, \& Margot(2012)]{fang2012} Fang, J., \& Margot, J.-L.\ 2012, \apj, 761, 92
\bibitem[Ford(2014)]{ford2014} Ford, E.~B.\ 2014, Proceedings of the National Academy of Science, 111, 12616
\bibitem[Forgan(2014)]{for14} Forgan, D. 2014 MNRAS, 437, 1352
\bibitem[Forte \& Mitrovica(1997)]{forte}Forte, A.M. \& Mitrovica, J.X., 1997, \nat, 390, 6661, p.676.
\bibitem[France et al.(2017)]{fra17} France, K., Fleming, B., West, G., et al. 2017, SPIE, 10397, 13
\bibitem[Gardner et al.(2006)]{JWST} Gardner, J.~P., Mather, J.~C., Clampin, M., et al.\ 2006, \ssr, 123, 485
\bibitem[Georgakarakos et al.(2018)]{georg18} Georgakarakos, N., Eggl, S., \& Dobbs-Dixon, I.\ 2018, \apj, 856, 155
\bibitem[Gildor and Tziperman, 2000]{gildor} Gildor, H. \& Tziperman, E., \ 2000, Paleoceanography and Paleoclimatology, 15, 605 
\bibitem[Gillon et al.(2017)]{Trapp} Gillon, M., Triaud, A.~H.~M.~J., Demory, B.-O., et al.\ 2017, \nat, 542, 456 
\bibitem[Gilmore, \& Ross(2008)]{Gil08} Gilmore, J.~B., \& Ross, A.\ 2008, \prd, 78, 124021.
\bibitem[Gilmozzi, \& Spyromilio(2007)]{EELT} Gilmozzi, R., \& Spyromilio, J.\ 2007, The Messenger, 127, 11
\bibitem[Gomes et al.(2005)]{LHB1} Gomes, R., Levison, H.~F., Tsiganis, K., \& Morbidelli, A.\ 2005, \nat, 435, 466 
\bibitem[Grazier(2016)]{Grazier} Grazier, K.~R.\ 2016, Astrobiology, 16, 23 
\bibitem[Grazier et al.(2018)]{Grazier2} Grazier, K.~R., Castillo-Rogez, J.~C., \& Horner, J.\ 2018, \aj, 156, 232.
\bibitem[Grazier et al.(2019)]{Grazier3} Grazier, K.~R., Horner, J., \& Castillo-Rogez, J.~C.\ 2019, \mnras, 490, 4388
\bibitem[Haghighipour \& Kaltenegger(2013)]{hag13} Haghighippour, N., Kaltenegger, L. 2013, ApJ, 777, 166
\bibitem[Hansen et al.(2013)]{hansen2013}Hansen, J., Sato, M., Russell, G. \& Kharecha, P., \ 2013. Philosophical Transactions of the Royal Society, 371(2001), 20120294.
\bibitem[Hatzes(2016)]{hatzes2016} Hatzes, A.~P.\ 2016, \ssr, 205, 267
\bibitem[Harakawa et al.(2015)]{Har15} Harakawa, H., Sato, B., Omiya, M., et al.\ 2015, \apj, 806, 5
\bibitem[Hays et al.(1976)]{hays} Hays, J.~D., Imbrie, J., \& Shackleton, N.~J.\ 1976, Science, 194, 1121
\bibitem[Hellier et al.(2011)]{HotJ2} Hellier, C., Anderson, D.~R., Collier-Cameron, A., et al.\ 2011, \apjl, 730, L31 
\bibitem[Herbert \& Fischer(1986)]{herb86} Herbert, T.D. \& Fischer, A.G. \ 1986, Nature, 321(6072), p.739.
\bibitem[Hewitt(2000)]{hew00} Hewitt, G. M., 2000, Nature, 405, 907
\bibitem[Hewitt(2004)]{hew04} Hewitt, G. M., 2004, Philosophical Transactions of the Royal Society B, 359, 1442
\bibitem[Hoffman et al., 1998]{Hoffman} Hoffman, P.F., Kaufman, A.J., Halverson, G.P. \& Schrag, D.P. \ 1998, Science, 281, 1342
\bibitem[Horner et al.(2003)]{Horner03} Horner, J., Evans, N.~W., Bailey, M.~E., et al.\ 2003, \mnras, 343, 1057
\bibitem[Horner \& Jones(2008)]{FoF1} Horner, J., \& Jones, B.~W.\ 2008, International Journal of Astrobiology, 7, 251 
\bibitem[Horner \& Jones(2009)]{FoF2} Horner, J., \& Jones, B.~W.\ 2009, International Journal of Astrobiology, 8, 75 
\bibitem[Horner et al.(2010)]{FoF3} Horner, J., Jones, B.~W., \& Chambers, J.\ 2010, International Journal of Astrobiology, 9, 1 
\bibitem[Horner, \& Lykawka(2010)]{QR322} Horner, J., \& Lykawka, P.~S.\ 2010, \mnras, 405, 49.
\bibitem[Horner \& Jones(2010)]{HabRev} Horner, J., \& Jones, B.W.\ 2010, International Journal of Astrobiology, 9, 273
\bibitem[Horner \& Jones(2012)]{FoF4} Horner, J., \& Jones, B.~W.\ 2012, International Journal of Astrobiology, 11, 147 
\bibitem[Horner et al.(2012)]{Anchises} Horner, J., M{\"u}ller, T.~G., \& Lykawka, P.~S.\ 2012, \mnras, 423, 2587.
\bibitem[Howard et al.(2012)]{SE4} Howard, A.~W., Marcy, G.~W., Bryson, S.~T., et al.\ 2012, \apjs, 201, 15
\bibitem[Huang \& Bani Shahabadi(2014)]{huang} Huang, Y., \& Bani Shahabadi, M.\ 2014, Journal of Geophysical Research (Atmospheres), 119, 13 
\bibitem[Huybers \& Wunsch(2005)]{huybers05}Huybers, P. \& Wunsch, C., 2005, \nat, 434(7032), p.491.
\bibitem[Imbrie et al.(1992)]{imbrie} Imbrie, John, E. A. Boyle, S. C. Clemens, A. Duffy, W. R. Howard, G. Kukla, J. Kutzbach et al. 1992, Paleoceanography, 7, 701
\bibitem[Ito et al.(1995)]{ito95}Ito, T., Masuda, K., Hamano, Y. \& Matsui, T., 1995, Journal of Geophysical Research: Solid Earth, 100, B8, pp.15147-15161.
\bibitem[Jansson \& Dynesius(2002)]{jansson} Jansson, R., \& Dynesius, M. 2002, Annu. Rev. Ecol. Evol. Syst., 33, 741
\bibitem[Johns et al.(2018)]{Johns18} Johns, D., Marti, C., Huff, M., et al.\ 2018, \apjs, 239, 14
\bibitem[Johns et al.(2012)]{GMT} Johns, M., McCarthy, P., Raybould, K., et al.\ 2012, \procspie, 8444, 84441H 
\bibitem[Kaltenegger et al.(2010)]{Kalt10} Kaltenegger, L., Selsis, F., Fridlund, M., et al.\ 2010, Astrobiology, 10, 89 
%\bibitem[Kane et al.(2007)]{kan07} Kane, S.R., Schneider, D.P., Ge, J. 2007, MNRAS, 377, 1610
\bibitem[Kane \& Gelino(2012)]{kan12} Kane, S.R., Gelino, D. 2012, AsBio, 12, 940
\bibitem[Kane \& Hinkel(2013)]{kan13} Kane, S.R., Hinkel, N.R. 2013, ApJ, 762, 7
\bibitem[Kane, \& Raymond(2014)]{kane2014} Kane, S.~R., \& Raymond, S.~N.\ 2014, \apj, 784, 104
\bibitem[Kane et al.(2016)]{kan16} Kane, S.R., Hill, M.L., Kasting, J.F., et al. 2016, ApJ, 830, 1
\bibitem[Kane \& Torres(2017)]{kan17} Kane, S.R., Torres, S.M. 2017, AJ, 154, 204
\bibitem[Kempton et al.(2018)]{kem18} Kempton, E.M.-R., Bean, J.L., Louie, D.R., et al. 2018, PASP, 130, 114401
\bibitem[Kinoshita(1975)]{kin75}Kinoshita, H., 1975, SAO Special report, 364.
\bibitem[Kirschvink et al.(2000)]{kirschvink} Kirschvink, J.L., Gaidos, E.J., Bertani, L.E. et al. 2000, PNAS, 4, 1400
\bibitem[Kirtland Turner(2014)]{KT2014} Kirtland Turner, S. 2014. Paleoceanography, 29(12), 1256-1266.
\bibitem[Kopparapu et al.(2013)]{kop13} Kopparapu, R.K., Ramirez, R.,
  Kasting, J.F., et al. 2013, ApJ, 765, 131
\bibitem[Kopparapu et al.(2014)]{kop14} Kopparapu, R.K., Ramirez,
  R.M., SchottelKotte, J., et al. 2014, ApJ, 787, L29
\bibitem[Lammer et al.(2009)]{Lammer09} Lammer, H., Bredeh{\"o}ft, J.~H., Coustenis, A., et al.\ 2009, \aapr, 17, 181 
\bibitem[Laskar et al.(1993)]{lask93} Laskar, J. Joutel, F. \& Robutel, P. 1993, \nat, 361, 615
\bibitem[Laskar et al.(2004)]{lask04} Laskar, J., Robutel, P., Joutel, F. et al. 2004, A\&A, 428, 261
\bibitem[Latham et al.(1989)]{Latham} Latham, D.~W., Mazeh, T., Stefanik, R.~P., Mayor, M., \& Burki, G.\ 1989, \nat, 339, 38 
\bibitem[Lewis et al.(2013)]{LewisFoF} Lewis, A.~R., Quinn, T., \& Kaib, N.~A.\ 2013, \aj, 146, 16 
\bibitem[Lingam \& Loeb(2018)]{Lingam18} Lingam, M., \& Loeb, A.\ 2018, \apjl, 857, L17
\bibitem[Linsenmeier et al.(2015)]{lins15}
Linsenmeier, M., Pascale, S. \& Lucarini, V., 2015, Planetary and Space Science, 105, pp.43-59.
\bibitem[Lisiecki \& Raymo(2005)]{lis2005} Lisiecki, L.E. \& Raymo, M.E. \ 2005, Paleoceanography, 20(1).
\bibitem[Lourens et al.(2005)]{lourens05}Lourens, L.J., Sluijs, A., Kroon, D., et al, 2005, \nat, 435, 7045, p.1083.
\bibitem[Lourens et al.(2010)]{lourens} Lourens, L.J., Becker, J., Bintanja, R. et al. 2010, Quat. Sci. Rev., 29, 352
\bibitem[MacDonald et al.(2016)]{Pack1} MacDonald, M.~G., Ragozzine, D., Fabrycky, D.~C., et al.\ 2016, \aj, 152, 105 
\bibitem[Marcy et al.(2014)]{dense} Marcy, G.~W., Isaacson, H., Howard, A.~W., et al.\ 2014, \apjs, 210, 20 
\bibitem[Maslin \& Ridgwell(2005)]{maslin} Maslin, M.A. \& Ridgwell, A.J.\ 2005, Geological Society, London, Special Publications, 247, 1
\bibitem[Masset \& Papaloizou(2003)]{Masset03} Masset, F.~S., \& Papaloizou, J.~C.~B.\ 2003, \apj, 588, 494 
\bibitem[Masuda(2014)]{Candy} Masuda, K.\ 2014, \apj, 783, 53 
\bibitem[Mayor \& Queloz(1995)]{51peg} Mayor, M., \& Queloz, D.\ 1995, \nat, 378, 355 
\bibitem[McGehee and Lehman, 2012]{mcgehee} McGehee, R. \& Lehman, C., 2012, SIAM Journal of Applied Dynamical Systems, 11, 684
\bibitem[Meadows et al.(2018)]{proxjwst} Meadows, V.~S., Arney, G.~N., Schwieterman, E.~W., et al.\ 2018, Astrobiology, 18, 133 
\bibitem[Milankovitch(1930)]{milank} Milankovitch, M. 1930, Mathematische Klimalehre und Astronomische Theorie der Klimaschwankungen. Handbuch der Klimatologie. 1 Teil A. von Gebrüder Borntraeger.
\bibitem[Mills et al.(2016)]{Pack2} Mills, S.~M., Fabrycky, D.~C., Migaszewski, C., et al.\ 2016, \nat, 533, 509 
\bibitem[Mojzsis et al.(2001)]{Mojzsis} Mojzsis, S.J., Harrison, T.M. \& Pidgeon, R.T. \ 2001, \nat, 409, 178
\bibitem[Morbidelli et al.(2010)]{LHB2} Morbidelli, A., Brasser, R., Gomes, R., Levison, H.~F., \& Tsiganis, K.\ 2010, \aj, 140, 1391 
\bibitem[Mullally et al.(2015)]{Mullaly15} Mullally, F., Coughlin, J.~L., Thompson, S.~E., et al.\ 2015, \apjs, 217, 31 
\bibitem[Nadachowska-Brzyska et al.(2015)]{Avian15} Nadachowska-Brzyska, K., Li, C., Smeds, L., Zhang, G. \& Ellegren, H., 2015, Current Biology, 25, 1375
\bibitem[Naish et al., 2009]{naish} Naish, T., Powell, R., Levy, R., et al., 2009, \nat, 458, 322
\bibitem[N{\'e}ron de Surgy and Laskar, 1997]{neron}N{\'e}ron de Surgy, O. \& Laskar,
J., Astronomy and Astrophysics, 318, pp.975-989
\bibitem[Nesvorn{\'y}(2018)]{nesvorny2018} Nesvorn{\'y}, D.\ 2018, Annual Review of Astronomy and Astrophysics, 56, 137
\bibitem[Nie et al.,(2008)]{nie08} Nie, J., King, J. and Fang, X. 2008, Geophysical Research Letters, 35, 21
%\bibitem[Norris et al.,(2002)]{NorrisBice} Norris, R.D., Bice, K.L., Magno, E.A. and Wilson, P.A. 2002 Geology, 30, 299
\bibitem[O'Malley-James et al.(2014)]{OMJ14} O'Malley-James, J.~T., Cockell, C.~S., Greaves, J.~S., \& Raven, J.~A.\ 2014, International Journal of Astrobiology, 13, 229 
\bibitem[Rauer et al.(2011)]{Rauer11} Rauer, H., Gebauer, S., Paris, P.~V., et al.\ 2011, \aap, 529, A8 
\bibitem[Raymo(1997)]{raymo97}Raymo, M.E., 1997, Paleoceanography, 12, 4, pp.577-585.
\bibitem[Raymo et al.(2006)]{raymo06}Raymo, M.E., Lisiecki, L.E. \& Nisancioglu, K.H., 2006, Science, 313, 5786, pp.492-495.
\bibitem[Raymond et al.(2018)]{raymond2018} Raymond, S.~N., Izidoro, A., \& Morbidelli, A.\ 2018, arXiv e-prints, arXiv:1812.01033
\bibitem[Royer et al.(2004)]{royer2004}
Royer, D.L., Berner, R.A., Montañez, I.P., Tabor, N.J. \$ Beerling, D.J., 2004. GSA today, 14(3), pp.4-10.
\bibitem[Ricker et al.(2015)]{ric15} Ricker, G.R., Winn, J.N.,
  Vanderspek, R., et al. 2015, JATIS, 1, 014003
\bibitem[Ridgwell et al.(1999)]{ridgwell} Ridgwell, A.J., Watson, A.J. and Raymo, M.E. 1999, Paleoceanography, 14, 437
\bibitem[Rubincam(1990)]{rub90}Rubincam, D.P., 1990, Science, 248, 4956, pp.720-721.
\bibitem[Saltzman \& Sutera(1987)]{saltz}Saltzman, B. \& Sutera, A., 1987. Journal of the Atmospheric Sciences, 44, 1, pp.236-241
\bibitem[Schwieterman et al.(2016)]{schjwst} Schwieterman, E.~W., Meadows, V.~S., Domagal-Goldman, S.~D., et al.\ 2016, \apjl, 819, L13 
\bibitem[Segura et al.(2005)]{Seg05} Segura, A., Kasting, J.~F., Meadows, V., et al.\ 2005, Astrobiology, 5, 706 
\bibitem[Sinukoff et al.(2016)]{SE5} Sinukoff, E., Howard, A.~W., Petigura, E.~A., et al.\ 2016, \apj, 827, 78 
\bibitem[Sniderman et al.(2013)]{snid13} Sniderman, J. M. K., Jordan, G. J. \& Cowling, R. M., 2013, Proceedings of the National Academy of Sciences of the United States of America, 110, 3423
\bibitem[Strom, \& Sprague(2003)]{Strom} Strom, R.~G., \& Sprague, A.~L.\ 2003, Exploring Mercury / Robert G. Strom and Ann L. Sprague. Springer-Praxis Books in Astronomy and Space Sciences. London (UK): Springer.\bibitem[Sullivan et al.(2015)]{sul15} Sullivan, P.W., Winn, J.N.,
  Berta-Thompson, Z.K., et al. 2015, ApJ, 809, 77
\bibitem[Tamuz et al.(2008)]{Tamuz08} Tamuz, O., S{\'e}gransan, D., Udry, S., et al.\ 2008, \aap, 480, L33 
\bibitem[Tarduno et al.(1998)]{Tarduno} Tarduno, J.A., Brinkman, D.B., Renne, P.R., Cottrell, R.D., Scher, H. and Castillo, P. 1998, Science, 282, 2241
\bibitem[Thorsett et al.(1993)]{psr1620} Thorsett, S.~E., Arzoumanian, Z., \& Taylor, J.~H.\ 1993, \apjl, 412, L33
\bibitem[Tsiganis et al.(2005)]{tsiganis2005} Tsiganis, K., Gomes, R., Morbidelli, A., et al.\ 2005, \nat, 435, 459
\bibitem[Turnbull \& Tarter(2003)]{Turnbull03} Turnbull, M.~C., \& Tarter, J.~C.\ 2003, \apjs, 145, 181 
\bibitem[Van Dam et al.(2006)]{vandam} Van Dam, J.A., Aziz, H.A., Sierra, M.Á.Á., Hilgen, F.J. et al., 2006, \nat, 443, 687
\bibitem[Vogt et al.(2010)]{SE3} Vogt, S.~S., Wittenmyer, R.~A., Butler, R.~P., et al.\ 2010, \apj, 708, 1366 
\bibitem[Volland(1996)]{voll96}Volland, H., 1996, Surveys in geophysics, 17, 1, pp.101-144.
\bibitem[Walker et al.(1981)]{walker} Walker, J.C., Hays, P.B. and Kasting, J.F.\ 1981, Journal of Geophysical Research: Oceans, 86, 9776
\bibitem[Walsh et al.(2012)]{LHB3} Walsh, K.~J., Morbidelli, A., Raymond, S.~N., O'Brien, D.~P., \& Mandell, A.~M.\ 2012, Meteoritics and Planetary Science, 47, 1941 
%\bibitem[Waltham(2006)]{walthamaxis} Waltham, D.\ 2006, International Journal of Astrobiology, 5, 327
\bibitem[Waltham(2019)]{waltearth} Waltham, D. \ 2019, Earth-Science Reviews, 192, 445
\bibitem[Ward et al.(1979)]{ward79}Ward, W.R., Burns, J.A. \& Toon, O.B., 1979, Journal of Geophysical Research: Solid Earth, 84, B1, pp.243-259.
\bibitem[Ward, \& Brownlee(2000)]{rareearth} Ward, P., \& Brownlee, D.\ 2000,  Rare earth : why complex life is uncommon in the universe / Peter Ward, Donald Brownlee. New York : Copernicus, c2000
\bibitem[Way \& Georgakarakos(2017)]{way17} Way, M.J., Georgakarakos, N. 2017, ApJ, 835, L1
\bibitem[Williams et al.(1993)]{wil93} Williams, M. A. J., Dunkerley, D. L., De Deckker, P., Kershaw, A. P. \& Stokes, T. J., 1993, Quaternary environments. In Quaternary environments; Edward Arnold Publishers Ltd.
%\bibitem[Williams(1994)]{wil94} Williams, G. 1994, AJ, 108, 711
\bibitem[Williams \& Pollard(2002)]{wil02} Williams, D.M., Pollard, D. 2002, IJAsB, 1, 61
\bibitem[Winn et al.(2011)]{SE2} Winn, J.~N., Matthews, J.~M., Dawson, R.~I., et al.\ 2011, \apjl, 737, L18 
\bibitem[Winn \& Fabrycky(2015)]{win15} Winn, J.N., Fabrycky, D.C. 2015, ARA\&A, 53, 409
\bibitem[Wittenmyer et al.(2007)]{Witt07} Wittenmyer, R.~A., Endl, M., Cochran, W.~D., \& Levison, H.~F.\ 2007, \aj, 134, 1276
\bibitem[Wittenmyer et al.(2011)]{etaEarth} Wittenmyer, R.~A., Tinney, C.~G., Butler, R.~P., et al.\ 2011, \apj, 738, 81 
\bibitem[Wittenmyer et al.(2012)]{24Sex} Wittenmyer, R.~A., Horner, J., \& Tinney, C.~G.\ 2012, \apj, 761, 165.
\bibitem[Wittenmyer et al.(2017)]{30177} Wittenmyer, R.~A., Horner, J., Mengel, M.~W., et al.\ 2017, \aj, 153, 167.
\bibitem[Wittenmyer et al.(2017)]{HD76920} Wittenmyer, R.~A., Jones, M.~I., Horner, J., et al.\ 2017, \aj, 154, 274.
\bibitem[Wright et al.(2012)]{HotJ1} Wright, J.~T., Marcy, G.~W., Howard, A.~W., et al.\ 2012, \apj, 753, 160 
\bibitem[Wolszczan \& Frail(1992)]{psr1257} Wolszczan, A., \& Frail, D.~A.\ 1992, \nat, 355, 145 
\bibitem[Wunsch(2003)]{wunsch2003}Wunsch, C., 2003, Climate Dynamics, 20, 4, pp.353-363.
\bibitem[Zachos et al.(2001)]{zachos} Zachos, J., Pagani, M., Sloan, L., Thomas, E. \& Billups, K. \ 2001, Science, 292, 686
\bibitem[Zeebe et al.(2017)]{zeebe}Zeebe, R. E., Westerhold, T., Littler, K. \& Zachos, J. C. \ 2017, Paleoceanography, 32, 440


\end{thebibliography}
\end{document}